\documentclass[useAMS,usenatbib]{mn2e}
\usepackage{epsfig}

\newcommand{\hd}{\rm{HD}}
\newcommand{\DD}{\rm{D_{2}}}
\newcommand{\Dm}{\rm{D}^{-}}
\newcommand{\Dp}{\rm{D}^{+}}
\newcommand{\mD}{\rm{D}}
\newcommand{\ddp}{{\rm D_{2}^{+}}}
\newcommand{\hdp}{{\rm HD^{+}}}
\newcommand{\me}{{\rm e^{-}}}
\newcommand{\mH}{{\rm H}}
\newcommand{\He}{{\rm He}}
\newcommand{\Hep}{{\rm He^{+}}}
\newcommand{\Hepp}{{\rm He^{++}}}
\newcommand{\Hp}{{\rm H}^{+}}
\newcommand{\Hm}{{\rm H}^{-}}
\newcommand{\mHtp}{{\rm H_{2}^{+}}}
\newcommand{\htp}{{\rm H_{3}^{+}}}
\newcommand{\mHt}{{\rm H_{2}}}
\newcommand{\hii}{{\rm H\,{\sc ii}} }
\newcommand{\expf}[3]{\exp \left(#1\frac{#2}{#3}\right)}
\def\simless{\mathbin{\lower 3pt\hbox
   {$\rlap{\raise 5pt\hbox{$\char'074$}}\mathchar"7218$}}}
\def\simgreat{\mathbin{\lower 3pt\hbox
   {$\rlap{\raise 5pt\hbox{$\char'076$}}\mathchar"7218$}}}

\title[Uncertainties in $H_{2}$ and HD Chemistry and
Cooling]{Uncertainties in $\mathbf{H_{2}}$ and HD Chemistry and Cooling and their
  Role in Early Structure Formation}
\author[S. C. O. Glover \& T. Abel]{S. C. O. Glover$^{1}$\thanks{E-mail: 
sglover@aip.de} \& T. Abel$^{2}$ \\ 
$^{1}$Astrophysikalisches Institut Potsdam, An der Sternwarte 16, 
D-14482 Potsdam, Germany \\
$^{2}$ Kavli Institute for Particle Astrophysics and Cosmology, Stanford University, 
Menlo Park, CA 94025, USA}

\begin{document}

\maketitle

\begin{abstract}
   At low temperatures, the main coolant in primordial gas is molecular
   hydrogen, $\mHt$. Recent work has shown that primordial gas that is
   not collapsing gravitationally but is cooling from an initially
   ionized state forms hydrogen deuteride, HD, in sufficient amounts to
   cool the gas to the temperature of the cosmic microwave
   background. This extra cooling can reduce the characteristic mass
   for gravitational fragmentation and may cause a shift in the
   characteristic masses of population III stars. Motivated by the
   importance of the atomic and molecular data for the cosmological
   question, we assess several chemical and radiative processes that
   have hitherto been neglected: the sensitivity of the low temperature
   $\mHt$ cooling rate to the ratio of ortho-$\mHt$ to para-$\mHt$, the
   uncertainty in the low temperature cooling rate of $\mHt$ excited by
   collisions with atomic hydrogen, the effects of cooling from $\mHt$
   excited by collisions with protons and electrons, and the large
   uncertainties in the rates of several of the reactions responsible
   for determining the $\mHt$ fraction in the gas.

   It is shown that the most important of neglected processes is the
   excitation of $\mHt$ by collisions with protons and electrons. This
   cools the gas more rapidly at early times, and
   so it forms less $\mHt$ and HD at late times. This fact, as
   well as several of the chemical uncertainties presented here,
   significantly affects the thermal evolution of the gas. We anticipate
   that this may lead to clear differences in future detailed three
   dimensional studies of first structure formation. In such
   calculations it has previously been shown that the details of the
   timing between cooling and merger events decides between immediate
   runaway gravitational collapse and a slower collapse delayed by
   turbulent heating.

   Finally, we show that although the thermal evolution of the gas is
   in principle sensitive to the ortho-para ratio, in practice the standard
   assumption of a 3:1 ratio produces results that are almost
   indistinguishable from those produced by a more detailed treatment.
\end{abstract}

\section{Introduction}
The very first stars to form in the Universe are believed to have formed within 
small protogalactic objects, cooled primarily by molecular hydrogen ($\mHt$).
Molecular hydrogen cooling becomes ineffective at temperatures below 
$T \simless 200 \: {\rm K}$, and at gas number densities $n > 10^{4} \: 
{\rm cm^{-3}}$, and so any dense fragments that form in the cooling and 
collapsing gas have a characteristic mass of a few hundred solar masses,
set by the Jeans mass at this temperature and density (\citealt{abn02}; 
\citealt{bcl02}; see also the reviews of \citealt{bl04} and \citealt{glo05}).
Since there is little evidence for sub-fragmentation during later stages of
the collapse \citep[although for a dissenting view see][]{cgk08}, and
since the high gas temperature leads to a high protostellar accretion
rate, there seems little to limit the growth of the first stars, which may
easily grow to $\sim 100 \: {\rm M_{\odot}}$ or more \citep[see e.g.][]{yoha06,oshn07}.

Efficient cooling from hydrogen deuteride, HD, can alter this scenario. HD
can cool the gas to lower temperatures than $\mHt$, and remains an
effective coolant up to higher densities, $n \sim 10^{6} \: {\rm cm^{-3}}$.
The characteristic mass of stars formed in HD-cooled gas is therefore
believed to be smaller, $\sim 10 \: {\rm M_{\odot}}$ \citep{jb06,yokh07},
reflecting the smaller characteristic mass scale imprinted on the 
cooling gas. However, HD cooling will only bring about a change of 
this kind in the characteristic mass scale if enough forms to cool the 
gas efficiently. \citet{bcl02} show that in simulations following the
formation of the very first stars, in protogalaxies with virial temperatures
$T_{\rm vir} < 10^{4} \: {\rm K}$, this does not occur: the inclusion of
deuterium chemistry and HD cooling has very little effect on the 
outcome. On the other hand, various authors have shown that in gas
cooling from an initially ionized state, enough HD forms to cool the 
gas down to temperatures close to the temperature of the cosmic 
microwave background \citep{nu02,no05,jb06,sv06,jgb07,yokh07}.
Note, however, that even without HD cooling the characteristic masses of
objects collapsing from gas within a relic primordial \hii
region have already been demonstrated to be smaller \citep{oshawn05}.

This difference in thermal evolution, depending on whether or not the gas
was once ionized, is a consequence of  the chemistry of HD formation
and destruction. The dominant reactions regulating the amount of
HD in the gas are
\begin{equation}
\mHt + \Dp \rightarrow \hd + \Hp,
\end{equation}
and 
\begin{equation}
\hd + \Hp \rightarrow \mHt + \Dp.
\end{equation}
Reaction 1 is exothermic, while reaction 2 is endothermic by 
0.0398~eV (462~K),
and so at low temperatures, chemical fractionation occurs: the
HD:$\mHt$ ratio becomes enhanced over the cosmological D:H
ratio by a large numerical factor. Consequently, even though 
the HD cooling rate per molecule decreases with decreasing 
temperature, the HD cooling rate per unit volume can actually 
increase, owing to the increase in the HD abundance produced
by this fractionation process \citep[see e.g.][]{glo07}. In 
conventional population III star formation calculations 
\citep[e.g.][]{abn02}, the fractional ionization is small, and
because of $pdV$ heating the gas temperature 
never becomes low enough for chemical fractionation to become  
efficient. Therefore, HD cooling remains unimportant. In contrast, in gas
cooling from an initially ionized state, more $\mHt$ forms, 
owing to the non-equilibrium fractional ionization in the cooling
gas \citep{sk87}, and the gas can reach a lower temperature. 
In practice, the extra cooling provided by the enhanced $\mHt$
abundance is sufficient to cool the gas to a point at which 
chemical fractionation becomes very important, following which
HD dominates the cooling.

Several processes and rate uncertainties, hitherto neglected, may
interfere with this simple picture. First, most calculations assume a
ratio of ortho-hydrogen ($\mHt$ with nuclear spin quantum number
$I=1$) to para-hydrogen ($\mHt$ with $I=0$) that is $(2I_{\rm ortho} +
1) / (2I_{\rm para} + 1) = 3$. This value is appropriate for warm
$\mHt$ in local thermodynamic equilibrium (LTE), which has many
different rotational and vibrational levels populated, but at low
temperatures and low densities, the ortho-para ratio may differ
significantly from this value. For instance, if only the $J=0$ and
$J=1$ rotational levels of the vibrational ground state are populated,
then the equilibrium ortho-para ratio is $9 \exp\left(-170.5 / T
\right)$. The relevance of this to the current situation lies in the
fact that the energy associated with the $v=0, J = 2 \rightarrow 0$
rotational transition in para-hydrogen, $E_{20} = 509.85 \: {\rm K}$,
is significantly smaller than the energy associated with the $v=0, J =
3 \rightarrow 1$ transition in ortho-hydrogen, $E_{31} = 844.65 \:
{\rm K}$. Consequently, para-hydrogen can cool the gas to lower
temperatures than ortho-hydrogen. It is therefore possible that the
ability of the gas to cool to the low temperatures required for HD
cooling to take over and dominate will be sensitive to the assumed
ortho-para ratio, and that the outcome of calculations that determine
it accurately will differ from that of calculations that assume a
ratio of 3:1.

A second issue affecting existing calculations is the fact that
the low temperature behaviour of the $\mHt$ cooling rate 
remains uncertain. The root cause of this uncertainty is the
sensitivity of the low energy $\mH$-$\mHt$ excitation 
cross-sections to the choice of the interaction potential 
used to calculate them. Most previous studies of HD cooling 
in primordial gas have made use of the fit to the low-density 
$\mHt$ cooling rate given by \citet{GP98}. At $T < 600 \: {\rm K}$, 
this fit is based on excitation rates from \citet{fbdl97} that were 
calculated using the BKMP2 potential energy surface of \citet{bkmp96}. 
However, recently \citet{wf07} have published a new set of $\mHt$ 
collisional excitation rate coefficients based on calculations performed
using the \citet{mgp02} potential energy surface. The $\mHt$ cooling 
function derived from these revised excitation rates differs
significantly from the \citet{GP98} rate at temperatures 
$T < 1000 \: {\rm K}$, but the consequences of this reduction in the 
cooling rate have yet to be explored in much detail.

A third issue regarding the $\mHt$ cooling rate is that fact that
most previous calculations have only included the effects of
collisional excitation of $\mHt$ by atomic hydrogen. However,
$\mHt$ can also be excited by collisions with $\mHt$, $\He$,
$\Hp$ and $\me$. As we show in \S\ref{res-cool}, in the conditions of
interest for HD formation, several of these neglected processes
play important roles.

The final issue affecting studies of the role of HD cooling 
that we examine here is the impact of the large uncertainties
that exist in several key rate coefficients for chemical reactions
involved in the formation and destruction of $\mHt$.
Although some of these uncertainties (which are discussed in 
detail in \S\ref{chem-model}) have received previous study in the 
literature \citep{SAV04,gsj06}, their impact on the ability of the gas
to cool to temperatures at which HD cooling becomes
dominant has not previously been explored.

In this paper, we explore these issues with the aid of
a detailed chemical and thermal model of primordial gas, 
coupled to two simple dynamical models. Our main aim is
to determine whether any of these sources of uncertainty can 
plausibly lead to significant differences in the evolution of the
gas, or whether existing results on the role of HD cooling are
robust. The structure of this paper is as follows. In 
Section~\ref{num}, we outline the numerical model used in
this work. In this context, we also discuss in more detail the
major uncertainties highlighted above. In Section~\ref{results},
we present and discuss our results, and we conclude in 
\S\ref{concl} with a brief summary. 

\section{Numerical model}
\label{num}
\subsection{Chemical network}
\label{chem-model}
To model the chemistry of $\mHt$ and HD in primordial gas, we use a chemical
network consisting of 115 reactions between 16 species, as summarized in 
Table~\ref{chem_gas}. This network differs significantly from previous
treatments of primordial deuterium chemistry in that it includes the formation and
destruction of doubly-deuterated hydrogen, ${\rm D_{2}}$. This is included 
because it has been suggested (D.~Savin, private communication) that conversion
of HD to ${\rm D_{2}}$ at low gas temperatures may be a significant destruction
mechanism for HD, although in practice we find that it is unimportant. 

For simplicity, we omit $\htp$, ${\rm HeH^{+}}$ and their deuterated analogues from
our chemical model. The abundances of these species are very small and their
influence on the cooling of the gas at intermediate to low densities is minimal
\citep{gs06,gs08}, so their omission should not significantly affect our results. We 
also omit lithium, for similar reasons.

We assume that any radiation backgrounds are negligible and so do not include
any processes involving photoionization or photodissociation. We also neglect
cosmic ray ionization; the influence of this latter process on promoting HD cooling
has been treated in detail elsewhere \citep{sv04,vs06,jce07,sb07}.

Whenever possible, rates for deuterated analogues of the basic hydrogen reactions
have been taken from the primary literature, or from the compilations of  \citet{sld98}, 
\citet{ws02} and \citet{wfp04}. However, some reactions do not appear to have been
previously considered in the astrochemical literature. In cases where we have been 
unable to find an appropriate rate, we have generally adopted the same procedure 
as in \citet{sld98}: for a non-deuterated reaction with a reaction rate that has a 
power-law temperature dependence $k \propto T^{m}$, we have generated the 
rates of the deuterated analogues by multiplying this rate by a scaling factor 
$(\mu_{\rm H} / \mu_{\rm D})^{m}$, where $\mu_{\rm H}$ and $\mu_{\rm D}$ 
are the reduced masses of the reactants in the non-deuterated and deuterated 
reactions respectively.

For reactions where the presence of a deuteron increases the number of distinguishable 
outcomes -- e.g.\ the dissociative attachment of $\hd$ with $\me$ (reactions 57--58), which
can produce either $\mH$ and $\Dm$ or $\Hm$ and $\mD$, in contrast to 
the dissociative attachment of $\mHt$ with $\me$ (reaction 23) which can only produce 
$\Hm$ and $\mH$ -- and where no good information exists on the branching ratio of the
reaction, we assume that the probability of each outcome is uniform. For this particular
example, this gives branching ratios of 50\% for reactions 57 and 58 respectively.

Finally, the rate coefficients for several of the included reactions require more detailed
discussion, which can be found in sections \ref{admn-rates}--\ref{hd-coll-dissoc} below.

\subsubsection{Associative detachment and mutual neutralization of $H^{-}$
\label{admn-rates}
and $D^{-}$}
The rates of reactions 2 \& 5, i.e. the associative detachment of $\Hm$ with $\mH$:
\begin{equation}
\Hm + \mH \rightarrow \mHt + \me,
\end{equation}
and the mutual neutralization of $\Hm$ with $\Hp$,
\begin{equation}
\Hm + \Hp \rightarrow \mH + \mH,
\end{equation}
are uncertain by up to an order of magnitude. When the fractional ionization of the gas
is small, these uncertainties are unimportant, as in this case reaction 2 proceeds much
faster than reaction 5. However, in gas with a high fractional ionization, such as gas
recombining from an initially ionized state, reaction 5 competes with reaction 2 for the
available $\Hm$ ions and so the uncertainties in the rates of these reactions introduce a significant
uncertainty into the amount of $\mHt$ that is formed. A large associative detachment rate 
and small mutual neutralization rate lead to the production of a larger $\mHt$ fraction 
(at a given time) than a small associative detachment rate and large mutual neutralization 
rate \citep{gsj06}. 

The default value for $k_{2}$ in our models is 
\begin{equation}
k_{2} = 1.3 \times 10^{-9} \: {\rm cm^{3}} \: {\rm s^{-1}},
\end{equation}
based on the measurement of \citet{sff67}. However, in \S\ref{res-chem} we present
results from models performed using 
\begin{equation}
k_{2} = 5.0 \times 10^{-9} \: {\rm cm^{3}} \: {\rm s^{-1}}
\end{equation}
and 
\begin{equation}
k_{2} = 0.65 \times 10^{-9} \: {\rm cm^{3}} \: {\rm s^{-1}},
\end{equation}
which represent plausible upper and lower bounds on the actual rate \citep{gsj06}.

Similarly, our default value for $k_{5}$ is given by 
\begin{equation}
k_{5} = 2.4 \times 10^{-6}  T^{-0.5} \left(1.0 + \frac{T}{20000} \right) \:  {\rm cm^{3}} \: {\rm s^{-1}},
\end{equation}
taken from \citet{cdg99},
but in \S\ref{res-chem} we also examine models using 
\begin{eqnarray}
k_{5} & = & 5.7 \times 10^{-6} T^{-0.5} + 6.3 \times 10^{-8} - 9.2 \times 10^{-11} T^{0.5} \nonumber \\
& & \mbox{} + 4.4 \times 10^{-13} T \:  {\rm cm^{3}} \: {\rm s^{-1}},
\end{eqnarray}
taken from \citet{MOS70} and
\begin{equation}
k_{5} = 7.0 \times 10^{-7} T^{-0.5}  \:  {\rm cm^{3}} \: {\rm s^{-1}},
\end{equation}
taken from \citet{dl87}. \citet{gsj06} have suggested that the last of these rates may be
erroneously small, owing to typographical errors in \citet{dl87}. Nevertheless, this 
rate has been used in a number of recent models of HD formation in primordial gas
\citep[see e.g.][]{no05,jb06}, justifying its consideration here.

In view of the large uncertainties in the rates of reactions 2 \& 5, we have assumed that identical 
rates apply for the deuterated analogues of these reactions (nos.\ 54--56 and 66--68), 
since any small differences in the basic rates caused by the presence of one or two 
deuterons in place of protons are likely swamped by this basic uncertainty.

\subsubsection{Charge transfer from $H^{+}$ to $H_{2}$ (reaction 7)}
\label{h2ct-rates}
The most accurate cross-section for this process at astrophysically
relevant energies is that computed by \citet{k02}; the corresponding thermal
rate coefficient is given in \citet{SAV04}. However, as \citet{SAV04}
discuss in some detail, a large number of other rates for this reaction
are given in the literature, differing by orders of magnitude at temperatures
below $10^{4} \: {\rm K}$. As this reaction is an important $\mHt$
destruction mechanism, particularly in gas recombining from an initially
ionized state, and as most previous studies of $\hd$ formation
in primordial gas have used one or another of these less accurate
rate coefficients (\citealt{yokh07} are a notable exception), it seems
appropriate to examine the effect that the choice of this rate coefficient 
has on the final amount of $\mHt$ formed and on the ability of the gas
to cool to temperatures at which HD cooling becomes dominant.
Therefore, while we use the \citet{SAV04} rate in most of our models,
we examine in \S\ref{res-chem} the effect of using two other rates from 
the literature.

The first of these, from \citet{sk87} 
\begin{equation}
k_{7} = 2.4 \times 10^{-9} \expf{-}{21200}{T} \: {\rm cm^{3}} \: {\rm s^{-1}}
\end{equation}
is, strictly  speaking, only applicable to vibrationally excited $\mHt$,
but in spite of this \citet{jb06} use this rate for charge transfer with
ground-state $\mHt$ in their study of $\hd$ cooling. As the 
comparison in Figure~1 of \citet{SAV04} demonstrates, this rate is
significantly larger than other determinations in the literature.

At the other extreme, \citet{ABE97} quote a rate
\begin{eqnarray}
k_{7} & = & \exp \left(-24.2491469  \right. \nonumber \\
& &  \mbox{}+ 3.4008244 (\ln T_{\rm e}) \nonumber \\
&  &   \mbox{}     - 3.8980040  (\ln T_{\rm e})^{2} \nonumber \\
& & \mbox{} + 2.0455878  (\ln T_{\rm e})^{3} \nonumber \\
&  &     \mbox{}     - 5.4161829 \times 10^{-1}  (\ln T_{\rm e})^{4} \nonumber \\
& &   \mbox{} + 8.4107750 \times 10^{-2}  (\ln T_{\rm e})^{5} \nonumber \\
&  &     \mbox{}     - 7.8790262 \times 10^{-3} (\ln T_{\rm e})^{6} \nonumber \\
&  &   \mbox{}         + 4.1383984 \times 10^{-4} (\ln T_{\rm e})^{7} \nonumber \\
&  &   \left. \mbox{}      - 9.3634588 \times 10^{-6} (\ln T_{\rm e})^{8} \right) \: {\rm cm^{3}} \: {\rm s^{-1}},
\end{eqnarray}
where $T_{\rm e}$ is the gas temperature in units of electron-volts. This
rate is based on \citet{JAN87}, and has subsequently been adopted by
a number of authors \citep[see e.g.][]{no05}. However, it is much smaller
at $T < 10^{4} \: {\rm K}$ than any of the other determinations in the \citet{SAV04}
comparison.

\subsubsection{Collisional dissociation of $H_{2}$ (reactions 8--11)}
In Table~\ref{chem_gas}, we list two rates for each process: one for $\mHt$ that is all in
the vibrational ground-state (appropriate for low density gas), and one
for $\mHt$ with local thermodynamic equilibrium (LTE) level populations.
At intermediate densities, we adopt a rate coefficient for each reaction given by
\begin{equation}
\log k_{\rm i}  = \left( \frac{n/n_{\rm cr}}{1 + n/n_{\rm cr}} \right)
\log k_{\rm i, LTE} + \left(\frac{1}{1 + n/n_{\rm cr}} \right) \log k_{\rm i, v=0},
\end{equation}
where $k_{\rm i}$ is the collisional dissociation rate for collisions with species
$i$, $k_{\rm v=0, i}$ and $k_{\rm LTE, i}$ are the rates for this reaction in
the $v=0$ and LTE limits respectively, and $n_{\rm cr}$ is the critical
density, given by
\begin{equation}
\frac{1}{n_{\rm cr}} = \frac{x_{\mH}}{n_{\rm cr, \mH}} + 
\frac{x_{\mHt}}{n_{\rm cr, \mHt}} + \frac{x_{\He}}{n_{\rm cr, \He}}.
\end{equation}
Here, $x_{\rm H} = n_{\mH} / n$, $x_{\mHt} = 2n_{\mHt} / n$, $x_{\He} = n_{\He} / n$, $n$ is
the number density of hydrogen nuclei, and
\begin{equation}
n_{\rm cr, \mH} =   {\rm dex} \left[3.0 - 0.416 \log{T_{4}} - 0.327 \left(\log{T_{4}} \right)^{2} \right], 
\end{equation}
\begin{equation}
n_{\rm cr, \mHt}  =  {\rm dex} \left[4.845 - 1.3 \log{T_{4}} + 1.62 \left(\log{T_{4}} \right)^{2} \right], 
\end{equation}
and
\begin{equation}
n_{\rm cr, \He} = {\rm dex} \left[5.0792 \left\{1.0 - 1.23 \times 10^{-5} (T - 2000) \right\} \right],
\end{equation}
with $T_{4} = T / 10000 \: {\rm K}$. The expression for $n_{\rm cr, \mH}$ is from 
\citet{ls83}, but has been decreased by an order of magnitude, as recommended 
by \citet{MAR96}. The expression for  $n_{\rm cr, \mHt}$ comes from \citet{sk87},
and the expression for $n_{\rm cr, \He}$ comes from \citet{drcm87}. Note that this 
expression for the critical density assumes that in high density gas, 
$n_{\rm e} \ll n_{\mH}$, so that electron excitation of $\mHt$ does not
significantly affect the value of $n_{\rm cr}$.

\subsubsection{${He^{+}}$ recombination (reaction 19)}
In optically thick gas that is a mixture of neutral H and He, the effective 
$\Hep$ recombination coefficient is given by
\begin{equation}
k_{19} = 0.68 k_{\rm 19, rr, A} + 0.32 k_{\rm 19, rr, B} 
+ k_{\rm 19, di},
\end{equation}
where we have assumed that $n_{\mHt} \ll n_{\mH}$; see \citet{ost89}
for a more detailed discussion.

In these conditions, it is also necessary to take account of the photoionization of
$\mH$ caused by the $\Hep$ recombination emission.  As long as the gas
is highly optically thick above the Lyman limit, this can be modelled as
a local $\mH$ ionization rate with a value
\begin{equation}
R_{\rm pi} = k_{\rm pi} n_{\rm e} n_{\Hep} \: {\rm cm^{-3}} \: {\rm s^{-1}},
\end{equation}
where
\begin{equation}
k_{\rm pi} = \left[0.68 k_{\rm 19, rr, A} + 0.28 k_{\rm 19, rr, B} 
+ k_{\rm 19, di} \right].
\end{equation}
We have not included a similar contribution from $\Hepp$ recombination, as
in conditions where the $\Hepp$ abundance is significant, we expect $\mH$
to be almost completely ionized.

\subsubsection{Three-body $H_{2}$ formation (reactions 30 \& 31)}
\label{h2tb-rates}
At high densities, reactions 30 \& 31 are important sources of $\mHt$.
However, the rate coefficients for these reactions are highly uncertain, as
previously discussed in \citet{glo07}. To assess the importance of this
uncertainty on our results, we have carried out simulations using two different values
for $k_{30}$: the first, taken from \citet{abn02} and partially based on
\citet{orel87} is the lowest of the values we have found in the literature:
\begin{equation}
k_{30} = \left \{ \begin{array}{lr}
1.14 \times 10^{-31} T^{-0.38} \: {\rm cm^{3}} \: {\rm s^{-1}} & \hspace{8pt} T \leq 300 \: {\rm K} \\
3.9 \times 10^{-30} T^{-1.0} \hspace{10pt} {\rm cm^{3}} \: {\rm s^{-1}} & \hspace{8pt} T > 300 \: {\rm K}
\end{array} \right.
\end{equation}
The other, taken from the recent paper of \citet{fh07} has the highest 
value at low temperatures of any of the rates we have found:
\begin{equation}
k_{30} = 1.44 \times 10^{-26} T^{-1.54} \: {\rm cm^{3}} \: {\rm s^{-1}}.
\end{equation}
To fix the rate of reaction 31, we follow \citet{pss83} and assume that
$k_{31} = k_{30} / 8$. 

\subsubsection{Destruction of ${D_{2}}$ by collision with H (reaction 107)}
The data tabulated in \citet{mie03} span the temperature range 
$200 \le T \le 2200 \: {\rm K}$. At lower temperatures, we simply
extrapolate our fit to the higher temperature data: this fit remains
well behaved at low temperatures, and since the rate of this reaction falls off 
exponentially at low $T$, we are not particularly sensitive to errors
in its value in this temperature range. At $T > 2200 \: {\rm K}$, we use
the simple exponential fit given by \citet{mie03} to their high temperature
calculations; although not formally valid at these temperatures, the fit
remains well-behaved, and hopefully lies not too far from the true value.

\subsubsection{Collisional dissociation of HD and $D_{2}$ (reactions 108--115)}
\label{hd-coll-dissoc}
For collisions with electrons, accurate rates are available in \citet{tt02a}
and \citet{tt02b}. For collisions with H, $\mHt$ or He, however, we have been
unable to find a treatment in the literature. We have therefore assumed that the
rates of these reactions in the $v=0$ and LTE limits are the same as for the
corresponding H reactions (nos.\ 8--10). For ${\rm D_{2}}$, we also adopt
the same value for the critical density, while for HD, we increase $n_{\rm cr}$
by a factor of 100 to account for its larger radiative transition probabilities. Note that
although these rates are highly approximate, this probably does not introduce
much uncertainty into the chemical model, as reactions 40 and 107 become
effective at much lower temperatures and therefore will generally dominate
the destruction of HD and ${\rm D_{2}}$ in warm gas.

\subsection{The ortho-para hydrogen ratio}
\label{op_rat_details}
In order to follow the evolution of the ortho-para hydrogen ratio in the gas,
we directly follow the time-dependent level populations of the lowest four
energy levels of the $\mHt$ molecule, the $J=0, 1, 2$ and 3 rotational levels 
of the vibrational ground state. Rates for collisional transitions between 
these four states are taken from several sources: non-reactive collisions
with $\mH$ (which cannot change the ortho-para ratio) are treated
using the rates computed by \citet{wf07}, while for reactive collisions
(which {\em can} change the ortho-para ratio), we use the rates suggested
by \citet{BOU99}. Collisions with protons are treated using the rates computed
by \citet{ger90}. Radiative transitions rates are taken from \citet{wsd98}.

Newly-formed $\mHt$ is assumed, for simplicity, to reside in the $J=0$
ground state. This assumption is not correct: $\mHt$ formed by associative 
detachment of $\Hm$ is, in general, highly excited and has 
a non-zero ortho-para ratio \citep{LAU91}. However, it is easy to show that
this assumption has little effect on the ortho-para ratio. In conditions 
where associative detachment dominates the destruction of $\Hm$, we 
can write the $\mHt$ formation timescale as
\begin{equation}
t_{\rm form} = \frac{x_{\mHt}}{k_{1} x_{\me} n},
\end{equation}
where $x_{\mHt}$ and $x_{\me}$ are the fractional abundances of $\mHt$
and free electrons respectively. In comparison, collisions with protons cause 
the ortho-para ratio to reach equilibrium on a timescale
\begin{equation}
t_{\rm op} = \frac{f_{\rm op}}{k_{\rm op} x_{\Hp} n},
\end{equation}
where $f_{\rm op}$ is the ortho-para ratio, $x_{\Hp}$ is the fractional
abundance of protons,  and $k_{\rm op}$ is an
appropriately averaged rate coefficient for the conversion of 
ortho-$\mHt$ to para-$\mHt$ by proton collision. From \citet{ger90},
we know that $k_{\rm op} \sim 10^{-10} \: {\rm cm^{3}} \: {\rm s^{-1}}$, while from
Table~\ref{chem_gas} we see that at a representative low temperature of 200~K,
$k_{1} = 2.07 \times 10^{-16} \: {\rm cm^{3}} \: {\rm s^{-1}}$. Thus, the timescales 
are comparable only if
\begin{equation}
x_{\mHt} \sim 10^{-6} f_{\rm op} \frac{x_{\me}}{x_{\Hp}}.
\end{equation}
If we make the reasonable assumption that $x_{\me} \simeq x_{\Hp}$,
and that $f_{\rm op}$ is of order unity, then this argument demonstrates 
that the $\mHt$ formation process has a significant effect on the ortho-para 
ratio only when the $\mHt$ fraction is very small, $x_{\mHt} \simless 10^{-6}$.

A similar comparison can also be performed between the $\mHt$ formation
timescale and the lifetimes of excited states of $\mHt$, but again the 
after-effects of the formation process are important only when the $\mHt$
fraction is very small.

Our model for the ortho-para ratio becomes increasingly inaccurate at high 
temperatures, as the excitation of states with $J > 3$ or $v > 0$ becomes 
important, but since the sensitivity of the $\mHt$ cooling rate to the 
ortho-para ratio is large only at low temperatures (see \S\ref{h2_cool_func} below), 
this simplified approach is sufficient for our purposes.

\subsection{Thermal processes}
\label{h2_cool_func}
\subsubsection{$H_{2}$ cooling: collisions with H}
As we have already discussed in Section 1, the low temperature rates for the
collisional excitation of $\mHt$ by $\mH$ are highly sensitive to the choice of
potential energy surface used to describe the ${\rm H_{3}}$ system
\citep{sd94}. An 
accurate determination of the $\mHt$ cooling function at low temperatures 
and low gas densities
requires a level of accuracy in the potential that has been difficult to achieve,
and as a consequence there are a number of determinations of the 
low-density limit of the $\mHt$ cooling function in the literature that differ 
substantially at  temperatures $T < 1000 \: {\rm K}$ \citep[see, for instance, 
the comparison in Figure~A1 of][]{GP98}. In recent years, the most widely 
used version has been that of \citet{GP98}:\footnote{This cooling rate has
units of ${\rm erg} \: {\rm cm^{3}} \: {\rm s^{-1}}$, as do all of the other cooling 
rates quoted in this paper, unless indicated otherwise.}
\begin{eqnarray}
\Lambda_{\rm \mHt, GP} &  = & {\rm dex} \left[ -103.0 + 97.59 \log T 
- 48.05 (\log T)^{2} \right. \nonumber \\
& & \left. \mbox{} + 10.80 (\log T)^{3} - 0.9032 (\log T)^{4} \right].
\end{eqnarray}
This rate is based on two separate sets of collisional rate coefficients. At
temperatures $T < 600 \: {\rm K}$, the rates used are those computed by 
\citet{fbdl97} using a fully quantal approach and the BKMP2 potential 
energy surface of \citet{bkmp96}. At $T > 600 \: {\rm K}$, the rates used
are those of \citet{mm93}, which were computed using the quasi-classical
trajectory approach and the LSTH potential energy surface 
\citep{l73,sl78,th78}. An ortho-para ratio of 3:1 is assumed at all 
temperatures.

Recently, however, \citet{wf07} have published a new set of collisional
rate coefficients computed using the potential energy surface of 
\citet{mgp02}. The rms error in this new potential energy surface
is more than an order of magnitude smaller than the error in the \citet{bkmp96}
potential, and \citet{wf07} argue that it should allow a more accurate
determination of the near-threshold behaviour of $\mHt$, and hence
a better determination of the low-temperature excitation rates and
cooling rate. \citet{wf07} and \citet{wgf07} show that there are indeed
significant differences in the low temperature behaviour of a number
of different excitation rates. 

We have used the rate coefficients calculated by \citet{wf07} to compute
separate cooling rates for ortho-$\mHt$ and para-$\mHt$ in the low density
limit due to collisions with atomic hydrogen. For ortho-$\mHt$, we 
find that the cooling rate in the temperature range $100 < T < 6000 \: 
{\rm K}$ is fit to within 2\% with a function of the form
\begin{equation}
\log \Lambda_{\rm H_{2}, H} = \sum_{i=0}^{7} a_{i} \log(T_{3})^{i},
\label{h2-fit}
\end{equation}
where $T_{3} = T / 1000 \: {\rm K}$. The fitting coefficients $a_{i}$ are listed in
Table~\ref{ortho-coeffs}. Below 100~K, we extrapolate the \citet{wf07} rate as
\begin{equation}
\Lambda_{\rm oH_{2}, H} = 5.09 \times 10^{-27} T_{3}^{1/2} \exp \left(\frac{-852.5}{T} \right).  
\end{equation}
Note that as HD cooling dominates at these low temperatures, we are not particularly
sensitive to errors in this extrapolation.

\begin{table}
\caption{Fitting coefficients for the cooling rate of ortho-$\mHt$ excited by collisions
with atomic hydrogen \label{ortho-coeffs}}
\begin{tabular}{crr}
\hline
Coefficient & $100 < T < 1000 \: {\rm K}$ & $ 1000 \le T < 6000 \: {\rm K}$ \\
\hline
$a_{0}$ & -24.330855 & -24.329086 \\
$a_{1}$ & 4.4404496 & 4.6105087 \\
$a_{2}$ & -4.0460989 & -3.9505350 \\
$a_{3}$ & -1.1390725 &  12.363818 \\
$a_{4}$ &  9.8094223 &  -32.403165 \\
$a_{5}$ & 8.6273872 & 48.853562  \\
$a_{6}$ & 0.0 & -38.542008 \\
$a_{7}$ & 0.0 &  12.066770 \\
\hline
\end{tabular}
\end{table}

For para-$\mHt$, we follow a similar procedure: the para-$\mHt$ cooling rate
for $100 < T < 6000 \: {\rm K}$ can again be fit to within 3\% by a function of the form of 
Equation~\ref{h2-fit}, using the fitting coefficients listed in Table~\ref{para-coeffs}.
At $T < 100 \: {\rm K}$, we use the extrapolation
\begin{equation}
\Lambda_{\rm pH_{2}, H} = 8.16 \times 10^{-26} T_{3}^{1/2} \exp \left(\frac{-509.85}{T} \right).  
\end{equation}

\begin{table}
\caption{Fitting coefficients for the cooling rate of para-$\mHt$ excited by collisions
with atomic hydrogen \label{para-coeffs}}
\begin{tabular}{crr}
\hline
Coefficient & $100 < T < 1000 \: {\rm K}$ & $ 1000 \le T < 6000 \: {\rm K}$ \\
\hline
$a_{0}$ & -24.216387 & -24.216387 \\
$a_{1}$ & 3.3237480 & 4.2046488 \\
$a_{2}$ & -11.642384 & -1.3155285 \\
$a_{3}$ & -35.553366  & -1.6552763 \\
$a_{4}$ & -35.105689  &  4.1780102 \\
$a_{5}$ & -10.922078 &  -0.56949697 \\
$a_{6}$ & 0.0 & -3.3824407 \\
$a_{7}$ & 0.0 &  1.0904027 \\
\hline
\end{tabular}
\end{table}

Given these partial rates, the total $\mHt$ cooling rate in the low-density limit
due to collisions with atomic hydrogen for gas with an ortho-hydrogen
abundance $x_{\rm o}$ and para-hydrogen abundance $x_{p}$ is then simply
\begin{equation}
\Lambda_{\rm H_{2}, H} =  \left(\frac{x_{o}}{x_{o} + x_{p}}\right) \Lambda_{\rm oH_{2}, H} + 
 \left(\frac{x_{p}}{x_{o} + x_{p}} \right) \Lambda_{\rm pH_{2}, H}. 
 \label{h2-op-lambda}
\end{equation}

\subsubsection{$H_{2}$ cooling: collisions with $H_{2}$}
To treat cooling due to collisions between two $\mHt$ molecules, we follow \citet{flpr00}
and use rates for the excitation of para-$\mHt$ and ortho-$\mHt$ by ground-state 
para-$\mHt$ derived from \citet{fr98} and rates for the excitation of para-$\mHt$ and 
ortho-$\mHt$ by ground-state ortho-$\mHt$ derived from \citet{fr99}. In the low density
limit, and for temperatures in the range $100 < T < 6000 \: {\rm K}$, these rates are fit to 
within 2\% by functions of the form
\begin{equation}
\log \Lambda_{\rm H_{2}, H_{2}} = \sum_{i=0}^{5} a_{i} \log(T_{3})^{i}.
\end{equation}
The fitting coefficients are listed in Tables~\ref{h2-coeffs-para} and \ref{h2-coeffs-ortho}
for cooling from para-$\mHt$ and ortho-$\mHt$ respectively. 

The total cooling rate in gas with an ortho-$\mHt$ abundance $x_{o}$ and a para-$\mHt$
abundance $x_{p}$ is given by
\begin{eqnarray}
\Lambda_{\mHt, \mHt} & = & x_{p}^{2} \Lambda_{\rm pH_{2}, pH_{2}} + x_{p} x_{o} 
\Lambda_{\rm pH_{2}, oH_{2}} \nonumber \\
& & \mbox{}  + x_{o} x_{p} \Lambda_{\rm oH_{2}, pH_{2}} 
+ x_{o}^{2} \Lambda_{\rm oH_{2}, oH_{2}} 
\end{eqnarray}
where $\Lambda_{\rm pH_{2}, pH_{2}}$ denotes the cooling rate due to the excitation
of para-$\mHt$ by para-$\mHt$, $\Lambda_{\rm pH_{2}, oH_{2}}$ the cooling rate due to 
the excitation of para-$\mHt$ by ortho-$\mHt$, etc.

\begin{table}
\caption{Fitting coefficients for the cooling rate of para-$\mHt$ excited by collisions
with $\mHt$ \label{h2-coeffs-para}}
\begin{tabular}{crr}
\hline
Coefficient & Para-$\mHt$ & Ortho-$\mHt$ \\
\hline
$a_{0}$ & -23.889798 & -23.748534 \\
$a_{1}$ & 1.8550774 & 1.76676480 \\
$a_{2}$ & -0.55593388 & -0.58634325 \\
$a_{3}$ & 0.28429361 & 0.31074159 \\
$a_{4}$ & -0.20581113 &  -0.17455629 \\
$a_{5}$ & 0.13112378 & 0.18530758 \\
\hline
\end{tabular}
\end{table}

\begin{table}
\caption{Fitting coefficients for the cooling rate of ortho-$\mHt$ excited by collisions
with $\mHt$ \label{h2-coeffs-ortho}}
\begin{tabular}{crr}
\hline
Coefficient & Para-$\mHt$ & Ortho-$\mHt$ \\
\hline
$a_{0}$ & -24.126177 & -24.020047 \\
$a_{1}$ & 2.3258217 &  2.2687566 \\
$a_{2}$ &  -1.0082491 & -1.0200304 \\
$a_{3}$ & 0.54823768 & 0.83561432 \\
$a_{4}$ & -0.33679759 &  -0.40772247 \\
$a_{5}$ & 0.20771406 & 0.096025713 \\
\hline
\end{tabular}
\end{table}

\subsubsection{$H_{2}$ cooling: collisions with He}
\label{h2he_rate}
The excitation of $\mHt$ by collisions with helium has been studied by a large 
number of authors \citep[see e.g.][and references therein]{lee05}. Currently, the 
most reliable theoretical calculations appear to be those performed using the 
\citet{mr94} ${\rm HeH_{2}}$ potential energy surface \citep[e.g.][]{frz98,bfd99,bvb99}. 
The more recent \citet{bmp03} surface, which was expected to be more accurate, 
produces results for some transitions that are in serious conflict with experimental 
determinations \citep{lee05} and so results derived using this potential energy 
surface are currently not considered reliable.

In our models, we use an $\mHt$ cooling rate due to collisions with $\He$ that is
derived from the calculations of \citet{frz98} for temperatures in the range 
$100 < T < 6000 \: {\rm K}$ and from \citet{bfd99} for temperatures $T < 100 \: {\rm K}$
(which were not treated in the Flower et~al.\ study). Comparison of the \citet{frz98} 
and \citet{bfd99} rates at temperatures $T > 100 \: {\rm K}$ shows that they agree
to within 10\%.  As before, we have derived separate rates for ortho-$\mHt$ and 
para-$\mHt$. In the low density limit, both cooling rates are fit to within 1\% for 
temperatures $T < 6000 \: {\rm K}$ by a function of the form
\begin{equation}
\log \Lambda_{\rm H_{2}, He} = \sum_{i=0}^{5} a_{i} \log(T_{3})^{i},
\end{equation}
where $T_{3} = T / 1000 \: {\rm K}$. The fitting coefficients $a_{i}$ for the
ortho and para cases are listed in Table~\ref{helium-coeffs}.

\begin{table}
\caption{Fitting coefficients for the cooling rate of $\mHt$ excited by collisions
with atomic helium \label{helium-coeffs}}
\begin{tabular}{crr}
\hline
Coefficient & Para-$\mHt$ & Ortho-$\mHt$ \\
\hline
$a_{0}$ & -23.489029 &  -23.7749  \\
$a_{1}$ &  1.8210825 &  2.40654 \\
$a_{2}$ & -0.59110559 & -1.23449 \\
$a_{3}$ &  0.42280623 &  0.739874 \\
$a_{4}$ & -0.30171138 &  -0.258940 \\
$a_{5}$ &  0.12872839 &   0.120573 \\
\hline
\end{tabular}
\end{table}

\subsubsection{$H_{2}$ cooling: collisions with protons and electrons}
In gas with a significant fractional ionization, collisions with protons and
electrons can lead to a substantial $\mHt$ cooling rate.  To treat the 
effect of collisions with protons, we use the rotational excitation rates of
\citet{ger90} and the vibrational cross-sections of \citet{k02}.
In the low density limit, the cooling rates of  ortho-$\mHt$
and para-$\mHt$ due to pure rotational transitions to levels with 
$2 \le J \le 7$ in the vibrational ground state, plus pure vibrational
transitions to levels with $1 \le v \le 4$ can be fit to within 2\% over the temperature 
range $10 < T < 10000 \: {\rm K}$ by a function of the form
\begin{equation}
\log \Lambda_{\rm H_{2}, \Hp} = \sum_{i=0}^{5} a_{i} \log(T_{3})^{i},
\end{equation}
where $T_{3} = T / 1000 \: {\rm K}$, using the fitting coefficients listed in 
Table~\ref{proton-coeffs}. Note that these cooling
rates include the effects of ortho-para interconversion in reactive 
collisions. In addition, it is also necessary to account for the effect
on the thermal balance of the gas of transitions from $J=0$ to $J=1$
and vice versa. Conversion of para-$\mHt$ to ortho-$\mHt$ in the
$J=0 \rightarrow 1$ transition cools the gas by $170.5 k \simeq 2.4 
\times 10^{-16}$~ergs per transition, while conversion of ortho-$\mHt$
to para-$\mHt$ in the $J=1 \rightarrow 0$ transition heats the gas
by the same amount. In thermodynamic equilibrium, the number of
transitions from $J=0$ to $J=1$ exactly balances the number of
transitions from $J=1$ to $J=0$, and so there is no net effect on the
gas temperature. However, if the gas is not in thermodynamic 
equilibrium, then there can be net heating or cooling of the gas,
depending upon whether the ortho-to-para ratio is greater than
or less than the equilibrium value. We account for this in our model
with a rate of the form
\begin{equation}
\Lambda_{\mHt, \Hp, 0 \leftrightarrow 1}   =   4.76 \times 10^{-24} 
\left[9 \exp\left(\frac{-170.5}{T}\right) x_{p} - x_{o} \right],  \label{base_rate}
\end{equation}
where $x_{o}$ and $x_{p}$ are the fractional abundances of ortho-$\mHt$
and para-$\mHt$, and where we have again made use of the rotational excitation 
and de-excitation rates of \citet{ger90}.

\begin{table}
\caption{Fitting coefficients for the cooling rate of $\mHt$ excited by collisions
with protons \label{proton-coeffs}}
\begin{tabular}{crr}
\hline
Coefficient & Para-$\mHt$ & Ortho-$\mHt$ \\
\hline
$a_{0}$ & -21.757160 & -21.706641   \\
$a_{1}$ & 1.3998367 & 1.3901283   \\
$a_{2}$ & -0.37209530  &  -0.34993699 \\
$a_{3}$ &  0.061554519 & 0.075402398 \\
$a_{4}$ & -0.37238286 & -0.23170723  \\
$a_{5}$ &  0.23314157 & 0.068938876  \\
\hline
\end{tabular}
\end{table}

We note that our treatment of $\mHt$ cooling due to collisions with $\Hp$ does
not account for the effects of rovibrational transitions, as the calculations by
\citet{k02} are not rotationally resolved. In view of the potential importance 
of this process in primordial gas cooling from an initially hot, ionized state, 
a more comprehensive treatment would be desirable.

To treat $\mHt$ excitation by collisions with free electrons, we use the
rates given by \citet{drd83}, based on cross-sections from \citet{ellt68},
\citet{cgm69} and \citet{ls71}. \citet{drd83} gives formulae
for the collisional de-excitation rates of pure rotational transitions with
$\Delta J = 2$ and pure vibrational transitions between $v =1, 2$ and 3
and the vibrational ground-state. Using these rates, we have computed
the low density para-$\mHt$--$\me$ and ortho-$\mHt$--$\me$ cooling
rates over a wide range of temperatures, and have fit them with functions
\begin{equation}
\log \Lambda_{\rm H_{2}, \me} = \log \left[\exp \left( \frac{-x}{kT} \right) \right] 
\times \sum_{i=0}^{5} a_{i} \log(T_{3})^{i}
\end{equation}
where $x = 509.85 k$ for para-$\mHt$ and $x = 845 k$ for ortho-$\mHt$.
The fitting coefficients are listed in Table~\ref{electron-coeffs}. In both cases, the fit is
accurate to within 10\% over the temperature range $10 < T < 10000 \: {\rm K}$. 
We note that as \citet{drd83} do not gives rates for rotational
transitions with $\Delta J > 2$ or for rovibrational transitions, our derived
$\mHt$ cooling rates will underestimate the true rates at high temperatures.
However, as collisions with protons are considerably more effective at exciting
$\mHt$ than collisions with electrons (see \S\ref{h2_tot_cool} below), the error that this
introduces into our calculations is unlikely to be large.

\begin{table}
\caption{Fitting coefficients for the cooling rate of $\mHt$ excited by collisions
with electrons \label{electron-coeffs}}
\begin{tabular}{crrr}
\hline
Coefficient & Para-$\mHt$ & Para-$\mHt$ &  Ortho-$\mHt$ \\
& $T \le 10^3 \: {\rm K}$ & $T > 10^3 \: {\rm K}$ & \\
\hline
$a_{0}$ & -22.817869 & -22.817869 & -21.703215  \\
$a_{1}$ & 0.95653474 &  0.66916141 & 0.76059565   \\
$a_{2}$ & 0.79283462  & 7.1191428 & 0.50644890  \\
$a_{3}$ & 0.56811779 &  -11.176835 & 0.050371349 \\
$a_{4}$ & 0.27895033 &  7.0467275 & -0.10372467   \\
$a_{5}$ & 0.056049813 & -1.6471816  & -0.035709409 \\
\hline
\end{tabular}
\end{table}

\subsubsection{$H_{2}$ cooling: the total cooling function}
\label{h2_tot_cool}
In the low density limit, the total cooling rate per $\mHt$ molecule (with
units of ${\rm erg} \: {\rm s^{-1}}$) is simply given by the sum of the 
cooling rates due to collisions with $\mH$, $\mHt$, $\He$, $\Hp$ and
$\me$, i.e.\
\begin{equation}
\Lambda_{\mHt, n \rightarrow 0}  = \sum_{k} \Lambda_{\mHt, k} n_{\rm k}
\end{equation}
where $k = \mH, \mHt, \He, \Hp, \me$.

At high densities, the $\mHt$ level populations are in local thermodynamic 
equilibrium, and the $\mHt$ cooling rate per molecule is independent of the 
chemical composition of the gas and is given by 
\begin{equation}
\Lambda_{\rm H_{2}, LTE} = \sum_{i, j>i} A_{ji} E_{ji} f_{j},
\end{equation}
where $A_{ij}$ is the radiative de-excitation rate for a transition from
level $j$ to level $i$, $E_{ji}$ is the corresponding energy, and $f_{j}$ 
is the fraction of $\mHt$ molecules in level $j$, computed assuming 
LTE. At intermediate densities, we follow \citet{GP98} and write the
$\mHt$ cooling rate as
\begin{equation}
\Lambda_{\mHt} = \frac{\Lambda_{\rm H_{2}, LTE}}{1 + \Lambda_{\rm H_{2}, LTE} /
\Lambda_{\mHt, n \rightarrow 0}}.
\end{equation}

\begin{figure}
\centering
\epsfig{figure=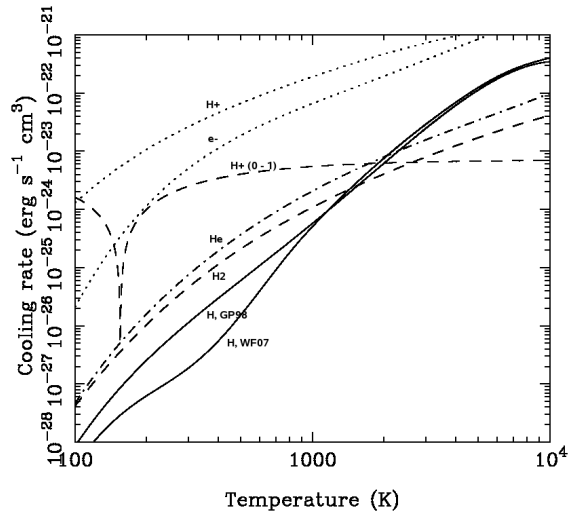,width=16pc,angle=-90}
\caption{$\mHt$ cooling rates per molecule, computed for $n = 10^{-4}
\: {\rm cm^{-3}}$, for
collisions with $\mH$ (lower solid line), $\mHt$ (lower dashed line), $\He$
(dash-dotted line), $\me$ (lower dotted line), and $\Hp$ (upper dotted
line; note that this rate excludes the effects of transitions between the 
ortho and para ground states).  In every case an ortho-para ratio of 3:1 is 
assumed. Transitions between the ortho and para ground states, brought about
by collisions with $\Hp$, cool the gas at $T \simgreat 150 \: {\rm K}$, and
heat it at lower temperatures; note, however, that the low-temperature heating
is a consequence of our adoption of a temperature-independent ortho-para 
ratio. Also shown is the widely used \citet{GP98} cooling function (upper 
solid line), which considers only collisions between $\mH$ and $\mHt$. 
\label{H2_cool_colliders}}
\end{figure}

In Figure~\ref{H2_cool_colliders}, we plot  $\Lambda_{\mHt, k}$ as a
function of temperature for  $k = \mH, \mHt, \He, \Hp$ and $\me$. In
this plot, we assume a fixed ortho-para ratio of 3:1 (corresponding to 
$x_{o} = 0.75$ and $x_{p} = 0.25$). We also include in this plot, for
the purposes of comparison, the widely used \citet{GP98} cooling
function.

It is immediately apparent from this plot that collisions between $\mH$
and $\mHt$ are relatively ineffective at cooling the gas at low 
temperatures. Given equal abundances of $\mH$ and $\mHt$, collisions
with $\mHt$ provide more cooling than collisions with $\mH$ for 
temperatures $T < 1400 \: {\rm K}$. Similarly, collisions with $\He$
provide more cooling than collisions with $\mH$ for $T < 2000 \: {\rm K}$,
while collisions with protons or electrons are more effective over the
whole of the temperature range examined here, again assuming equal
abundances. 

Of course, in reality, the abundances of the various collision partners will 
generally not be equal: in low density primordial gas, at the temperature 
of interest here, atomic hydrogen is by far the most abundant species. 
Typically, in gas undergoing gravitational collapse within a small protogalaxy
($T_{\rm vir} < 10000 \: {\rm K}$) forming in a region of the intergalactic
medium (IGM) not
yet affected by stellar feedback, one finds abundances relative to atomic
hydrogen of $x_{\He} = 0.0825$ for He, $x_{\mHt} \sim 10^{-3}$ for $\mHt$
and $x_{\Hp} \simeq x_{\me} \sim 10^{-4}$ for protons and electrons.
If these relative abundances are taken into account, then atomic hydrogen
becomes comparatively more effective. Collisions with $\mHt$ become
completely unimportant for the whole of the temperature range studied,
and collisions with electrons can also be neglected. However, collisions
with helium remain important at low temperatures, and in fact dominate
the $\mHt$ cooling rate for $T < 650 \: {\rm K}$, despite the significantly
larger abundance of hydrogen relative to helium. Collisions with protons
are also important at $T \simless 400 \: {\rm K}$, in spite of the low proton
abundance.

In gas cooling from an initially ionized state, similar conclusions hold
regarding the relative importance of collisions with $\mH$, $\mHt$ and
$\He$. However, in this case, values for $x_{\Hp}$ and $x_{\me}$ that
are 10--100 times larger are not uncommon,  and the effects of $\mHt$-$\Hp$ 
and $\mHt$-$\me$ collisions are therefore much greater. 

It is also interesting to compare the relative importance of the various
processes if one adopts the \citet{GP98} rate for cooling from 
$\mHt$-$\mH$ collisions in place of our value derived from \citet{wf07}.
Figure~\ref{H2_cool_colliders}
demonstrates that the \citeauthor{GP98} cooling rate provides 
significantly more cooling at $T < 1000 \: {\rm K}$ than the newer 
\citeauthor{wf07} cooling rate, with the rates differing most significantly at 
temperatures $300 \simless T \simless 500 \: {\rm K}$, where the \citeauthor{GP98} 
rate provides almost five times more cooling than the comparable \citeauthor{wf07}
rate. Because of this, collisions with $\He$ and with protons and electrons
are less effective in comparison to collisions with $\mH$ at $T < 1000 \: {\rm K}$
when one uses the \citeauthor{GP98} rate. Nevertheless, even though it is
no longer the dominant process, cooling from $\mHt$-$\He$ collisions remains 
important at low temperatures, as it can contribute 20--30\% of the total $\mHt$
cooling rate. Furthermore, $\mHt$-$\Hp$ and $\mHt$-$\me$ collisions will
also still be important if the fractional ionization of the gas is large ($x_{\Hp}
\simgreat 10^{-3}$).

We should note at this point that we are not the first authors to highlight 
the potential importance of $\mHt$-$\He$ and $\mHt$-$\Hp$ collisions 
for cooling primordial gas. \citet{BOU99} include the effects of $\mHt$-$\He$
collisions in their calculations of the $\mHt$ cooling function, as do 
\citet{ss06}; the importance of helium is also discussed at some length
in \citet{flpr00}. The possible importance of $\mHt$-$\Hp$ collisions
was noted by \citet{GP98} and their effects were examined in more detail 
by \citet{fp00} and \citet{flpr00}, although only pure rotational transitions
were considered. On the other hand, to the best of our knowledge, we 
are the first authors to consider the effects of $\mHt$-$\me$ collisions in
primordial gas.

Finally, although in this section we have given fits to the cooling rates of
ortho-$\mHt$ and para-$\mHt$ separately, since we are interested in the
effects of varying the ortho-para ratio, we recognize that for some purposes
it may be useful to have the rates for a gas that has the often-assumed
3:1 mix of ortho and para-$\mHt$. In Table~\ref{total-coeffs}, we list fits to 
the low density $\mHt$ cooling rates due to collisions with $\mH$, $\mHt$, 
$\He$, $\Hp$ and $\me$ for this case. All of these fits are of the form
\begin{equation}
\log \Lambda_{\rm H_{2}} = \sum_{i=0}^{5} a_{i} \log(T_{3})^{i},
\end{equation}
where $T_{3} = T / 1000 \: {\rm K}$
and the accuracies are comparable to the accuracies of the separate
ortho and para-$\mHt$ fits. Note that the $\mHt$-$\Hp$ rate quoted 
here does not include the effects of collisional transitions from 
$J=0$ to $J=1$ or vice versa. However, this can be included through
the use of Equation~\ref{base_rate} with $x_{p} = 0.25$ and $x_{o} = 0.75$
for the 3:1 ortho-para ratio case.

\begin{table}
\caption{Fitting coefficients for $\mHt$ cooling rates, for a 3:1 ortho-para ratio
\label{total-coeffs}}
\begin{tabular}{ccl}
\hline
Species & Temperature range (K) & Coefficients \\
\hline
H & $10 < T \le 100$ & $a_{0} = -16.818342$ \\
 & &  $a_{1} =  37.383713 $ \\
 & &   $a_{2} = 58.145166 $ \\
 & &   $a_{3} = 48.656103 $ \\
 & &   $a_{4} = 20.159831 $ \\
  & &   $a_{5} = 3.8479610 $ \\
  & & \\
H & $100 < T \le 1000$ & $a_{0} = -24.311209$ \\
& & $a_{1} = 3.5692468$ \\
& & $a_{2} =  -11.332860$ \\
& & $a_{3} = -27.850082$ \\
& & $a_{4} = -21.328264$ \\
& & $a_{5} = -4.2519023$ \\
& & \\
H & $1000 < T \le 6000$ &  $a_{0} = -24.311209$ \\
& & $a_{1} = 4.6450521$  \\
& & $a_{2} =  -3.7209846$ \\
& & $a_{3} = 5.9369081$ \\
& & $a_{4} = -5.5108047$ \\
& & $a_{5} =  1.5538288$ \\
& & \\
$\mHt$ & $100 < T \le 6000$ & $a_{0} =  -23.962112$ \\
& & $a_{1} = 2.09433740$ \\
& & $a_{2} =  -0.77151436$ \\
& & $a_{3} = 0.43693353$ \\
& & $a_{4} = -0.14913216$ \\
& & $a_{5} = -0.033638326$ \\
& & \\
$\He$ & $10 < T \le 6000$ & $a_{0} = -23.689237$ \\
& & $a_{1} =   2.1892372$  \\
& & $a_{2} =  -0.81520438$ \\
& & $a_{3} =  0.29036281$ \\
& & $a_{4} =  -0.16596184$ \\
& & $a_{5} =  0.19191375$ \\
& & \\
$\Hp$ & $10 < T \le 10000$ & $a_{0} = -21.716699$ \\
& & $a_{1} =  1.3865783$ \\
& & $a_{2} =  -0.37915285$ \\
& & $a_{3} =  0.11453688$ \\
& & $a_{4} = -0.23214154$ \\
& & $a_{5} = 0.058538864$ \\
& & \\
$\me$ & $10 < T \le 200$ & $a_{0} =  -34.286155$ \\
& & $a_{1} = -48.537163 $  \\
& & $a_{2} =  -77.121176$ \\
& & $a_{3} =  -51.352459$ \\
& & $a_{4} =  -15.169160 $ \\
& & $a_{5} = -0.98120322$ \\
& & \\
$\me$ & $200 < T \le 10000$ & $a_{0} = -22.190316$ \\
& & $a_{1} = 1.5728955$ \\
& & $a_{2} = -0.21335100$ \\
& & $a_{3} = 0.96149759$ \\
& & $a_{4} = -0.91023195$ \\
& & $a_{5} = 0.13749749$ \\
\hline
\end{tabular}
\end{table}

\subsubsection{$H_{2}$ cooling: sensitivity to the ortho-para ratio}
\begin{figure}
\centering
\epsfig{figure=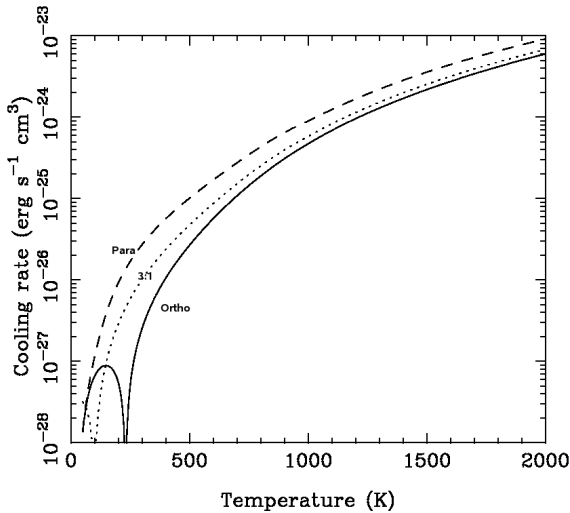,width=16pc,angle=-90}
\caption{Comparison of the $\mHt$ cooling rate per $\mHt$ molecule,
for gas with only ortho-$\mHt$ (solid line), para-$\mHt$ (dashed line) or a 3:1
mix of ortho and para-$\mHt$ (dotted line). We assume that $x_{\He} = 0.0825$, 
$x_{\mHt} = 0.001$ and $x_{\Hp} = x_{\me} = 10^{-4}$. Note that for
$T < 230 \: {\rm K}$ in the ortho-$\mHt$ case and $T < 98 \: {\rm K}$ in 
the 3:1 ratio case, the rate plotted is the net heating rate, after accounting
for heating due to the collisional conversion of $J=1$ ortho-$\mHt$ to
$J=0$ para-$\mHt$ by protons.
\label{H2_cool}}
\end{figure}

In Figure~\ref{H2_cool} we compare three different $\mHt$ cooling
rates: one for pure ortho-$\mHt$, one for pure para-$\mHt$, and one 
for which we assumed the standard 3:1 ortho-para ratio. In each
case, we assume that $n \ll n_{\rm crit}$, so that we are in the low
density limit, and adopt fractional abundances relative to hydrogen 
of $x_{\He} = 0.0825$, $x_{\mHt} = 0.001$, and $x_{\Hp} = x_{\me} = 
10^{-4}$ for He, $\mHt$, $\Hp$ and electrons respectively. Note that
at $T < 230 \: {\rm K}$ in the ortho-$\mHt$ case and for $T < 98 \:
{\rm K}$ in the 3:1 ratio case, collisional conversion of ortho-$\mHt$
in the $J=1$ rotational level to para-$\mHt$ in the $J=0$ rotational
level by protons heats the gas, and that the lowest temperature portions
 of the curves plotted in Figure~\ref{H2_cool} for these two cases therefore
represent {\em heating} rates.

The figure demonstrates the importance of the ortho-para $\mHt$ ratio
in determining the $\mHt$ cooling rate at temperatures below a few
hundred K. For instance, at $T = 300 \: {\rm K}$, there is a difference 
of an order of magnitude between the cooling rate of 
para-$\mHt$ and the cooling rate of ortho-$\mHt$. Because of this
large disparity, para-$\mHt$ will provide most of the contribution to
the $\mHt$ cooling rate at these temperatures even in gas that 
contains primarily ortho-$\mHt$; e.g.\ it is the 25\% of para-$\mHt$
that provides most of the cooling in the 3:1 case. Consequently,
relatively small deviations in the ortho-para ratio may have a large
effect on the low temperature $\mHt$ cooling rate. 

Furthermore, although the specific values for the cooling rates
plotted in Figure~\ref{H2_cool} are sensitive to our assumed 
chemical abundances, the basic point that the low temperature
$\mHt$ cooling rate is highly sensitive to the assumed ortho-para
ratio is robust, as it is a consequence of the difference in the energy
separations of the lowest levels of para-$\mHt$ ($E_{20} = 509.85 
\: {\rm K}$) and ortho-$\mHt$ ($E_{31} = 844.65 \: {\rm K}$).

\subsubsection{HD cooling}
To model $\hd$ cooling, we use the cooling function of \citet{lna05}. This
parameterization of the HD cooling rate assumes that HD-H collisions
make the dominant contribution. This is a much safer assumption in the
case of HD cooling than in the case of $\mHt$ cooling. Excitation rate
coefficients for HD-H collisions are typically much larger than for
$\mHt$-H collisions, and \citet{flpr00} show that they are comparable
to the excitation rates for HD-He or HD-$\mHt$ collisions. As 
$n_{\mH} \gg n_{\He} \gg n_{\mHt}$ in the conditions of interest here,
this means that the HD-H contribution will dominate.

The larger excitation rates for HD-H collisions also reduce the importance
of collisions with protons or electrons. Although accurate excitation rate 
coefficients for HD-$\Hp$ or HD-$\me$ collisions do not appear to be 
available, it seems reasonable to assume that they will be of a similar 
order of magnitude to the corresponding processes with $\mHt$. If so,
then in the $T < 200 \: {\rm K}$ temperature regime in which HD cooling
is important, collisions with electrons or protons become comparable
to collisions with atomic hydrogen only for fractional ionizations 
$x \simgreat 0.1$. To find such a large fractional ionization in gas this
cold would appear to be highly unlikely, and so it seems relatively safe
to neglect the effects of collisions with protons or electrons.

Although the \citet{lna05} parameterization of the HD cooling rate is
formally valid only in the temperature range $100 < T < 2 \times 10^{4} \: {\rm K}$, we have
compared its behaviour at lower temperatures with an explicit calculation of the 
cooling rate made using radiative de-excitation rates from \citet{arv82} and collisional rates
extrapolated from those computed by \citet{wgf07}. We find that the \citet{lna05} rate 
remains reasonably accurate down to temperatures as low as 50~K, with errors no 
greater than 20\%, and that even at $T = 30 \: {\rm K}$ it remains accurate to within 
a factor of two. At temperatures $T \gg 100 \: {\rm K}$, the \citet{lna05} cooling rate
slightly underestimates the effects of HD cooling compared to the newer calculations
of \citet{wgf07}, presumably owing to the more accurate vibrational excitation rates
used in the latter, but the differences are relatively small and in any case occur in the
temperature regime in which $\mHt$ cooling dominates. The breakdown of the 
\citet{lna05} fit at very high temperatures ($T > 20000 \: {\rm K}$) is unimportant, 
as the gas in our models never exceeds this temperature, nor could HD
cooling ever be significant in this temperature regime where the
cooling from the Lyman $\alpha$ line of neutral hydrogen peaks.

To correctly  model the effects of $\hd$ cooling at low temperatures, it is 
necessary to take the effects of the cosmic microwave background (CMB) into account. 
We do this approximately, by using a modified HD cooling rate, 
$\Lambda_{\rm HD}^{\prime}$, defined as
\begin{equation}
\Lambda_{\rm HD}^{\prime} = \Lambda_{\rm HD}(T) - \Lambda_{\rm HD}(T_{\rm CMB})
\end{equation}
where $\Lambda_{\rm HD}(T)$ and $ \Lambda_{\rm HD}(T_{\rm CMB})$ are the 
unmodified HD cooling rates at the gas temperature $T$ and the CMB temperature
$T_{\rm CMB}$ respectively.

The quoted range of densities for which the \citet{lna05} cooling function is valid
is $1 < n < 10^{8} \: {\rm cm^{-3}}$. To extend the range of the cooling function
to densities $n < 1 \: {\rm cm^{-3}}$, we assume that at these densities the cooling
rate per molecule is directly proportional to $n$,  and hence that
\begin{equation}
\Lambda_{\rm HD}(n = n^{\prime}) = n^{\prime} \Lambda_{\rm HD}(n=1)
\end{equation}
for $n^{\prime} \le 1 \: {\rm cm^{-3}}$, where $\Lambda_{\rm HD}(n)$ is the cooling
rate per HD molecule (with units ${\rm erg} \: {\rm s^{-1}}$) at gas number 
density $n$. To extend the cooling function to high densities, 
$n > 10^{8} \: {\rm cm^{-3}}$, we
assume that the HD molecule is in LTE and thus has a cooling rate per molecule
that is independent of density. In this regime,
\begin{equation}
\Lambda_{\rm HD}(n > 10^{8}) = \Lambda_{\rm HD}(n = 10^{8}).
\end{equation}
In view of the fact that $1 \ll n_{\rm cr, HD} \ll 10^{8} \: {\rm cm^{-3}}$, where 
$n_{\rm cr, HD}$ is the HD critical density, both of these assumptions appear
well justified.

\subsubsection{$H_{2}^{+}$, $HD^{+}$ and $D_{2}^{+}$ cooling}
In view of its possible importance in hot, ionized gas
\citep[see][]{yokh07}, we include the effects of vibrational
cooling from the $\mHtp$ molecular ion, as well as from
its deuterated analogues $\hdp$ and $\ddp$.

At low densities, the main contributions to the $\mHtp$ cooling rate come
from excitations by collisions with electrons and with neutral hydrogen
\citep{ss78}. We have computed the cooling rate due to collisions with
electrons, using the vibrational rates of \citet{st93} for excitations from
$v=0$ to $v=1$ and $v=2$; excitations to higher vibrational states 
($v=3$ to $v=8$) are also included, under the assumption that the
de-excitation rates for these transitions are comparable to the 
de-excitation rate from $v=2$ to $v=0$. We note that even at 
temperatures as high as $10^{4} \: {\rm K}$, at least half of the total
cooling comes from excitations to $v=1$ and $v=2$, and so our
approximate treatment of  transitions to the higher vibrational states
makes the cooling rate uncertain by at most a factor of a few at
high temperatures (and by far less than this at low temperatures).
The resulting $\mHtp$ cooling rate, $\Lambda_{\rm e, \mHtp}$,
 is given at $T \le 2000 \: {\rm K}$  by
\begin{equation}
\Lambda_{\rm e, \mHtp}  = 1.1 \times 10^{-19} T^{-0.34} \expf{-}{3025}{T},
\end{equation}
and at $T > 2000 \: {\rm K}$ by
\begin{equation}
\Lambda_{{\rm e}, \mHtp}  =  3.35 \times 10^{-21}  T^{0.12} \expf{-}{3025}{T}.
\end{equation}
For $\mHtp$ cooling arising from collisions with $\mH$, we use at $T \le 1000 \: {\rm K}$ a rate
\begin{equation}
\Lambda_{\mH, \mHtp}  = 1.36 \times 10^{-22}  \expf{-}{3152}{T},
\end{equation}
and at $T > 1000 \: {\rm K}$ a rate
\begin{equation}
\Lambda_{\mH, \mHtp} = {\rm dex} \left[ -36.42 + 5.95 \log(T) - 0.526 \log(T)^{2} \right].
\label{galli_fit}
\end{equation}
The high temperature rate is a fit made by \citet{GP98} to the rate given
in \citet{ss78}; note that owing to a normalization error, the rates given for 
$\Lambda_{\mH, \mHtp}$ and $\Lambda_{{\rm e}, \mHtp}$ in 
Figure~A2 of \citet{GP98} are too large by a factor of ten.
The low temperature rate given here is a physically reasonable extrapolation of
the \citet{ss78} rate that has the correct exponential fall-off at low
temperature.

At high densities, the vibrational levels of $\mHtp$ will be in LTE. In this
regime, the cooling rate per $\mHtp$ ion is given approximately by
\begin{equation}
\Lambda_{\rm LTE, \mHtp} = 2.0 \times 10^{-19} T^{0.1} \expf{-}{3125}{T}.
\end{equation}
To compute this rate, we included contributions from all vibrational
states $v \le 8$ and used level energies from \citet{kh07} and radiative transition
rates from \citet{pdp83}. The effects of rotational excitation were not included, but
are unlikely to change this expression by a large amount, owing to the
very small transition rates associated with these transitions.

At intermediate densities, we assume that the $\mHtp$ vibrational cooling
rate per $\mHtp$ ion is given approximately by the function
\begin{equation}
\Lambda_{\mHtp} = \frac{\Lambda_{\rm LTE, H_{2}^{+}}}{1 + 
\Lambda_{\rm LTE, H_{2}^{+}} / \Lambda_{\rm n \rightarrow 0, H_{2}^{+}}},
\label{h2p_full}
\end{equation}
where $\Lambda_{\rm n \rightarrow 0, H_{2}^{+}}$ is the cooling rate per
$\mHtp$ ion in the low density limit, and is given by
\begin{equation}
 \Lambda_{\rm n \rightarrow 0, H_{2}^{+}} = \Lambda_{\rm e, \mHtp} n_{\me} 
 + \Lambda_{\rm H, \mHtp} n_{\mH}.
\end{equation}

To model cooling from vibrational transitions in $\hdp$, we assume, in the absence
of better information, that the low density cooling rate is the same as that used for
$\mHtp$. However, since $\hdp$ has much larger radiative transition rates than
$\mHtp$, the LTE cooling rate for $\hdp$ is much larger than that for $\mHtp$.
We have calculated the $\hdp$ LTE cooling rate per ion using level energies from 
\citet{kh07} and transition rates from \citet{phb79}, and have fit it with the function
\begin{equation}
\Lambda_{\rm LTE, \hdp} = 1.09 \times 10^{-11} T^{0.03} \expf{-}{2750}{T}
\end{equation}
at temperatures $T \le 1000 \: {\rm K}$ and
\begin{equation}
\Lambda_{\rm LTE, \hdp} = 5.07 \times 10^{-12} T^{0.14} \expf{-}{2750}{T}
\end{equation}
at $T > 1000 \: {\rm K}$. For densities between the low density and LTE limits,
we use a function of the form of Equation~\ref{h2p_full} to compute the $\hdp$
cooling rate.

Finally, to model $\ddp$ cooling, we simply assume that the same rates apply 
as for $\mHtp$ cooling. In practice, the very small size of the typical $\ddp$ 
abundance renders this process irrelevant.

\subsubsection{Other processes}
In addition to the coolants listed above, we also include radiative cooling 
from the electronic excitation of H, He and $\Hep$ using rates taken from 
\citet{CEN92} and \citet{bbft00}, Compton cooling (again using a rate from
 \citealt{CEN92}), and bremsstrahlung (using the rates given in \citealt{sk87}).

Moreover, we also include the effects of chemical cooling from the collisional 
ionization of $\mH$, $\He$ and $\Hep$ (reactions 12, 17 and 18), collisional 
dissociation of  $\mHt$ (reactions 8--11), the destruction of $\mHt$ by charge 
transfer (reaction 7), and the recombination of $\Hp$, $\Hep$ and $\Hepp$ 
(reactions 13, 19 and 20), as well as chemical heating arising from the 
formation of $\mHt$ via reactions 2 and 4. 

Further details of our treatment of these processes can be found in \citet{gj07}.
 
\subsection{Model setup and initial conditions}
We model the chemical and thermal evolution of primordial gas within the
context of two simple toy models for its dynamical evolution. In one, we
assume that the gas evolution is isobaric.  In this model, the gas temperature
evolves as
\begin{equation}
\frac{{\rm d}T}{{\rm d}t} = \mbox{} - \frac{\gamma - 1}{\gamma}
\frac{\mu}{k} \frac{\Lambda - \Gamma}{\rho} + T \frac{{\rm d}\ln{\mu}}{{\rm d}t}
+ \frac{T}{\gamma-1} \frac{{\rm d}\ln{\gamma}}{{\rm d}t}
\end{equation}
where $\mu$ is the mean molecular weight of the gas (in grammes), $\Lambda$ 
and $\Gamma$ are the total cooling and heating rates per unit volume, and the
other symbols have their usual meanings. The rate of change of $\mu$ can 
be easily determined from the chemical rate equations. For the adiabatic index 
$\gamma$, we use the expression
\begin{equation}
\gamma = \frac{5 + 5 x_{\He} + 5 x_{\rm e} - 3 x_{\mHt}}{3 + 3 x_{\He} + 3 x_{\rm e} 
- x_{\mHt}},
\end{equation}
where $x_{\He}$, $x_{\mHt}$ and $x_{\rm e}$ are the fractional abundances
of helium, $\mHt$ and free electrons relative to the abundance of hydrogen
nuclei. In practice, $x_{\mHt}$ remains small at all densities encountered in 
our isobaric models, and so $\gamma \simeq 5/3$ throughout. Finally, at any
point in the evolution of the gas, we can relate the gas density to the temperature by
\begin{equation}
\rho = \left(\frac{T_{\rm i}}{T}\right) \left(\frac{\mu}{\mu_{\rm i}}\right) \rho_{\rm i}
\end{equation}
where $\rho_{\rm i}$, $T_{\rm i}$ and $\mu_{\rm i}$ are the initial values 
of the density, temperature and mean molecular weight respectively.

In the other model, we assume that the gas undergoes gravitational collapse 
at the free-fall rate. In this model, the gas temperature evolves as 
\begin{equation}
\frac{{\rm d}T}{{\rm d}t} = \frac{\gamma-1}{\rho} \left[ T \frac{{\rm d} \rho}{{\rm d} t} 
- \frac{\mu}{k} (\Lambda - \Gamma) \right] + \frac{T}{\gamma - 1} \frac{{\rm d}\gamma}{{\rm d}t} 
+ T \frac{{\rm d}\ln{\mu}}{{\rm d}t},
\end{equation}
and the gas density evolves as
\begin{equation}
 \frac{{\rm d} \rho}{{\rm d} t} = \frac{\rho}{t_{\rm ff}} 
\end{equation}
where $t_{\rm ff} = \sqrt{3\pi / 32 G \rho}$ is the free-fall time. 

The first of these models approximates the case of gas that has been strongly
shocked (e.g.\ by a supernova blast-wave or in a halo merger) but that
is not yet gravitationally unstable. The free-fall collapse model 
approximates the other extreme case, in which gas is highly gravitationally 
unstable, and contracts at the maximal rate. Realistically, the
dynamical evolution of gas involved in high-redshift structure formation
probably lies somewhere in between these two cases.

Although one could use far more sophisticated models for the dynamical
evolution of the gas \citep[see e.g.][]{yokh07}, our use of these simple models 
allows us to rapidly explore the effects
of the various different sources  of uncertainty discussed in this paper,
and to highlight which are deserving of more numerically expensive
three dimensional studies, and which
are unimportant and can be safely ignored in future work.

To evolve the coupled set of chemical rate equations and the thermal
energy equation we use the DVODE solver of \citet{bbh89}. 

We adopt standard helium and deuterium abundances of 
$x_{\He} = 0.0825$ and $x_{\mD} = 2.6 \times 10^{-5}$ relative to hydrogen
\citep{mol07}, and begin our simulations with gas in which hydrogen and 
deuterium are fully ionized and the helium is singly ionized. The initial 
abundances of all other species are set to zero. 

We fix the initial temperature at $T_{\rm i} = 20000 \: {\rm K}$, and examine
models with three different initial densities: $n_{\rm i} = 0.03$, 1, and $30
\: {\rm cm^{-3}}$. We run all of our models for two different redshifts, $z=10$
and $z=20$; the latter value is perhaps more appropriate for the study 
of the earliest generations of star formation, but the former allows us to
look at the effects of having a CMB temperature that is much smaller than
the temperature that the gas can reach through $\mHt$ cooling alone.

\section{Results}
\label{results}
\subsection{Ortho-para ratio}
\label{res-op}
In order to establish the effect that variations in the $\mHt$ ortho-para
ratio have on the thermal evolution of primordial gas, we considered
four separate cases: the two limiting cases in which all of the $\mHt$
is in the form of ortho- or para-hydrogen respectively, a third case in
which the standard ratio of 3:1 was assumed, and a final case in which
the ortho-para ratio was determined self-consistently, although approximately, 
from the populations of the lowest four rotational levels, as outlined
in \S\ref{op_rat_details}.

\begin{figure}
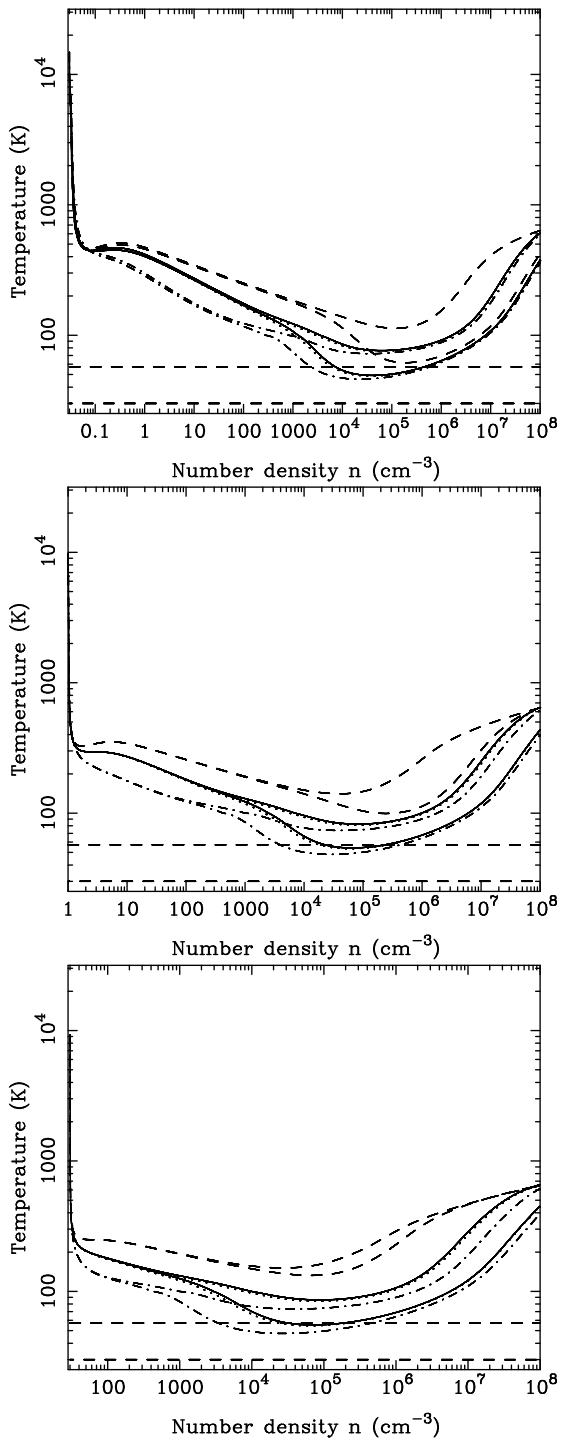

\centering
\epsfig{figure=f3a.eps,width=15pc,angle=-90}
\epsfig{figure=f3b.eps,width=15pc,angle=-90}
\epsfig{figure=f3c.eps,width=15pc,angle=-90}
\caption{(a) Temperature evolution as a function of gas number density
in free-fall collapse models with initial density $n_{\rm i} = 0.03 \: {\rm cm^{-3}}$
and initial redshifts $z=20$ (upper set of curves) and $z=10$ (lower set of curves).
Four cases are examined: gas with pure ortho-$\mHt$ (dashed curves), pure
para-$\mHt$ (dash-dotted curves), a 3:1 ortho-para ratio (solid curves) and an
ortho-para ratio determined by solution of the simplified level population
calculation discussed in \S\ref{op_rat_details} (dotted curves; note that these 
are barely distinguishable from the solid lines in the plot). The CMB
temperature at $z=10$ and $z=20$ is indicated by the horizontal dashed
lines.
(b) As (a), but for gas with $n_{\rm i} = 1 \: {\rm cm^{-3}}$.
(c) As (a), but for gas with $n_{\rm i} = 30 \: {\rm cm^{-3}}$. \label{ffcomp_op}}
\end{figure}

In Figure~\ref{ffcomp_op}a, we show how the gas temperature evolves
as a function of density for these four cases in two free-fall collapse
models with initial density $n_{\rm i} = 0.03 \: {\rm cm^{-3}}$ for redshifts
$z=10$ (lower set of curves) and $z=20$ (upper set of curves). 
Figures~\ref{ffcomp_op}b and \ref{ffcomp_op}c show similar results
for models with initial densities $n_{\rm i} = 1$ and $30 \: {\rm cm^{-3}}$
respectively. In each figure, the dashed and dash-dotted curves correspond
to calculations in which the $\mHt$ is all in ortho or para form respectively,
the solid curves correspond to the calculations that assume an ortho-para
ratio of 3:1 and the dotted curves correspond to the calculations in which
the ortho-para ratio was determined dynamically. The horizontal dashed
lines give the CMB temperature at $z=20$ (upper line) and $z=10$
(lower line).

Figure~\ref{ffcomp_op} demonstrates that there are significant differences
between the  temperature evolution in the pure ortho-$\mHt$, pure 
para-$\mHt$, and 3:1 ratio calculations. The para-$\mHt$ and 3:1 ratio
calculations differ primarily at $n < 10^{5} \: {\rm cm^{-3}}$, with temperatures
differing by as much as 50\% at $n \sim  100 \: {\rm cm^{-3}}$. At $n > 10^{5}
\: {\rm cm^{-3}}$, however, the simulations become convergent, and little difference
remains in the temperature evolution. The ortho-$\mHt$ simulations show an even
greater difference in behaviour. In most of these simulations, the gas temperature
remains significantly larger than in the para-$\mHt$ or 3:1 ratio runs, 
differing by a factor of two or more, and failing to converge with the other 
simulations even at $n > 10^{5} \: {\rm cm^{-3}}$. The gas temperature in most of
the ortho-$\mHt$ simulations remains above $100 \: {\rm K}$ throughout the collapse,
and although a comparison of cooling rates shows that HD cooling does
become dominant in these simulations, it does not succeed in driving down the
temperature to the same extent as in the other runs. The one exception
is the simulation with $n_{\rm I} = 0.03 \: {\rm cm^{-3}}$ and $z = 10$, which does
cool significantly below $100 \: {\rm K}$ and which converges with the
corresponding para-$\mHt$ and 3:1 ratio calculations.
 
\begin{figure}
\centering
\epsfig{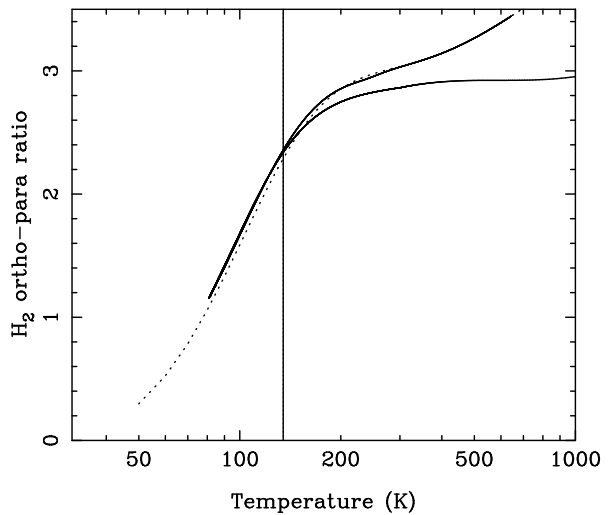}
\caption{Ortho-para ratio as a function of gas temperature in a free-fall
collapse model with $z=20$ and $n_{\rm i} = 1 \: {\rm cm^{-3}}$. The
collapsing gas evolves initially towards lower temperatures along the
lower branch of the solid line, before progressing back toward higher
temperatures along the upper branch as the gas heats up at high 
density. The dotted line gives the equilibrium ortho-para ratio for
our simplified model $\mHt$ molecule. The vertical solid line indicates
the temperature below which HD cooling becomes dominant. \label{ffop}}
\end{figure}

Despite the apparent sensitivity of the temperature evolution to the ortho-para
ratio, Figure~\ref{ffcomp_op} demonstrates that there is essentially no difference
between the results of calculations in which the ortho-para ratio is fixed at 3:1
or calculated self-consistently from the $\mHt$ level populations.  Figure~\ref{ffop}
helps to demonstrate why this is so. In the figure, we show the dependence
of the ortho-para ratio on the gas temperature in a representative free-fall
collapse model with $z=20$ and $n_{\rm i} = 1 \: {\rm cm^{-3}}$.  At 
temperatures $T > 200 \: {\rm K}$, the ortho-para ratio is approximately three,
both at low densities (solid line, lower branch) and at high densities (solid
line, upper branch). At lower temperatures, the ortho-para ratio falls off
steeply with decreasing temperature, and at the lowest temperature reached
by the gas, $T \simeq 81 \: {\rm K}$, para-hydrogen is almost as abundant as
ortho-hydrogen. However, in this calculation, HD cooling dominates over
$\mHt$ cooling at a temperature $T = 135 \: {\rm K}$ (indicated in the figure
by the vertical solid line). At this temperature the ortho-para ratio is $\sim 2.4$,
and so the $\mHt$ cooling rate does not differ greatly from the rate that
we obtain by assuming a fixed ortho-para ratio of 3:1. In other words, at the 
temperatures where the true $\mHt$ cooling rate differs significantly from the 
$\mHt$  cooling rate in the 3:1 ortho-para case, $\mHt$ cooling is itself 
unimportant,  and HD cooling dominates.  We have verified that the same 
explanation also serves to explain the results of our other free-fall collapse models.

\begin{figure}
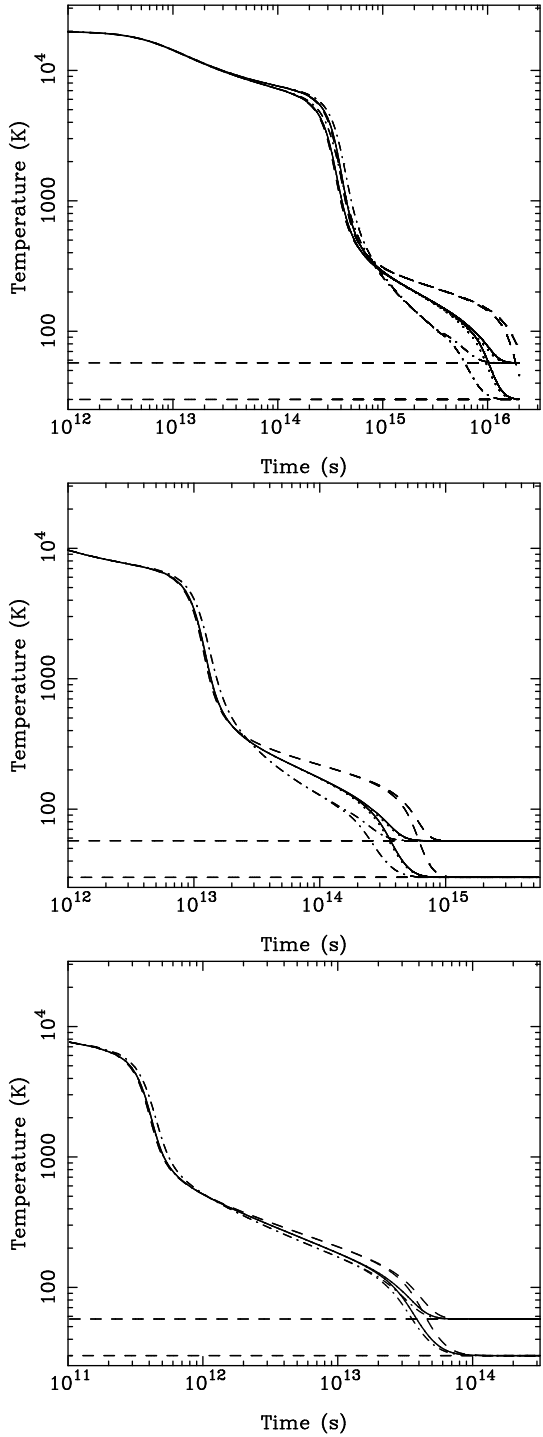

\centering
\epsfig{figure=f5a.eps,width=15pc,angle=-90}
\epsfig{figure=f5b.eps,width=15pc,angle=-90}
\epsfig{figure=f5c.eps,width=15pc,angle=-90}
\caption{(a) Temperature evolution as a function of time in
a set of models in which the gas evolution is isobaric. These
models have an initial gas number density $n_{\rm i} = 0.03 \: 
{\rm cm^{-3}}$, and results are plotted for both $z=10$ (lower
set of lines) and $z=20$ (upper set of lines).  Four cases are 
examined: gas with pure ortho-$\mHt$ (dashed lines), pure para-$\mHt$ 
(dash-dotted lines), a 3:1 ortho-para ratio (solid lines) and an ortho-para 
ratio determined from the level populations of our model $\mHt$
molecule (dotted lines;  note that again these are barely distinguishable 
from the solid lines in the plot). The CMB temperature at $z=10$ and $z=20$ 
is indicated by the horizontal dashed lines.
(b) As (a), but for an initial number density $n_{\rm i} = 1 \: {\rm cm^{-3}}$.
(c) As (a), but for an initial number density $n_{\rm i} = 30 \: {\rm cm^{-3}}$.
\label{isocomp_op}}
\end{figure}

Finally, to check that our conclusions do not depend on our choice of
dynamical model, we have examined the behaviour of isobarically
evolving gas in the same four cases, as illustrated in 
Figure~\ref{isocomp_op}. We see that again the temperature
evolution is sensitive to extreme variations of the ortho-para ratio,
but that the results of the self-consistent calculation are barely
distinguishable from those of the calculation assuming a 3:1 ortho-para 
ratio. Further investigation demonstrates that the reason for this
similarity is the same as in the free-fall collapse case: at temperatures
where $\mHt$ cooling is significant, the ortho-para ratio remains
close to three, while at the low temperatures at which it differs 
significantly from three, $\mHt$ cooling is unimportant and HD
cooling dominates.

We can therefore conclude that the adoption of a fixed ortho-para
ratio of 3:1, although strictly speaking unjustified at 
$T < 200 \: {\rm K}$, is nevertheless an adequate assumption  
for modelling the temperature evolution of primordial gas, 
and that therefore this potential source of uncertainty ultimately
proves to be unimportant.

\subsection{Choice of $\mathbf{H_{2}}$ cooling function}
\label{res-cool}
To explore the sensitivity of the thermal evolution of primordial gas, 
and in particular of the ability of the gas to cool to temperatures at 
which HD cooling dominates, to uncertainties and omissions in the
treatment of $\mHt$ cooling, we ran a number of free-fall collapse 
and isobaric evolution models using different $\mHt$ cooling 
functions. Our reference model (hereafter CF1)
uses the cooling function outlined in \S\ref{h2_cool_func} and used elsewhere in this paper;
it includes the effects of  collisions with $\mH$, $\mHt$, $\He$, $\Hp$ 
and $\me$, and an ortho-para ratio that was computed self-consistently
with the evolution of the gas. We also examined the effects of
omitting the $\Hp$ and $\me$ contributions (CF2), and of omitting
$\Hp$, $\me$ and $\He$ (CF3); note that in the latter case, the
$\mHt$ cooling function essentially consists only of the $\mHt$-$\mH$
contribution, as the $\mHt$ fractions in our calculations are too
small for $\mHt$-$\mHt$ collisions to ever become important.
Finally, we examine the effect of including only the $\mHt$-$\mH$
contribution, but using the \citet{GP98} cooling rate instead of
the \citet{wf07} rate (CF4). Note that case CF4 assumes a fixed 
ortho-para ratio of 3:1, while the other treatments determine the
ortho-para ratio self-consistently, as outlined above. However, the 
results of the previous section demonstrate that in practice this should 
not be a major source of error.

\begin{figure}
\centering
\epsfig{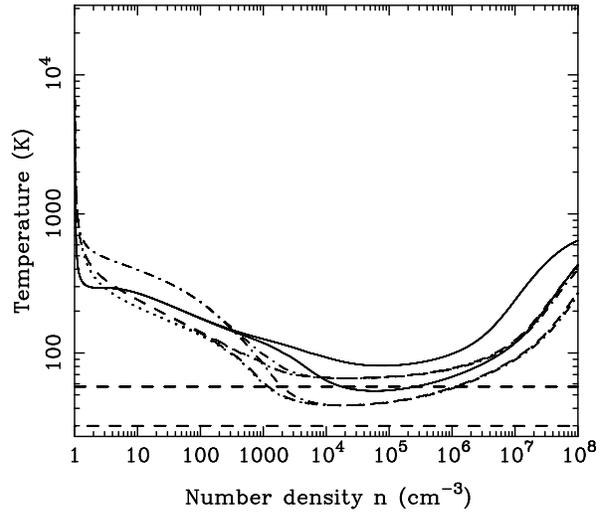}
\caption{Temperature evolution as a function of gas number density
in free-fall collapse models with initial density $n_{\rm i} = 1 \: {\rm cm^{-3}}$ 
and initial redshifts $z=10$ (upper curves) and $z=20$ (lower curves).
The  horizontal dashed lines indicate the CMB temperature at these
redshifts. Four different treatments of $\mHt$ cooling are compared:
our full model, CF1 (solid line); two variants of this model, one which omits
$\Hp$ and $\me$ collisions (CF2; dashed line), and one which omits
$\Hp$, $\me$ and $\He$ collisions (CF3; dot-dashed line); and the
treatment used in most previous studies, which includes only collisions
with $\mH$, but uses the \citet{GP98} cooling rate rather than the
\citet{wf07} rate. \label{ff_cool}}
\end{figure}

In Figure~\ref{ff_cool}, we show the temperature evolution of the gas
as a function of density in two free-fall collapse models with an
initial density $n_{\rm i} = 1 \: {\rm cm^{-3}}$ and redshifts $z=10$
and $z=20$. The solid, dashed, dot-dashed and dotted lines 
correspond to CF1, CF2, CF3 and CF4 respectively. There are
two important points to note about this plot.

First, it is clear that the temperature evolution of the gas in case CF1 
differs from that in the other models over the whole range of density 
studied here. Although the initial cooling of the gas is rapid in this model,
this only lasts until the temperature reaches $T \sim 300 \: {\rm K}$.
At lower temperatures, the cooling of the gas slows down 
dramatically, allowing cooling in the other models to catch up and
surpass it. The gas reaches a minimum temperature of $81 \: {\rm K}$
in the $z=20$ run and $53 \: {\rm K}$ in the $z=10$ run, significantly
higher than the limits set by the CMB. 

Second, the temperature evolution of the gas in models CF2, CF3 and
CF4 differs noticeably at densities $n < 10^{4} \: {\rm cm^{-3}}$ and 
temperatures $T < 500 \: {\rm K}$. The behaviour of models CF2 and
CF4 is surprisingly similar, given the difference in physical content
of these two models, but the behaviour of model CF3 is clearly 
different, with the gas in the latter model cooling less rapidly than in 
the other two. At densities higher than $10^{4} \: {\rm cm^{-3}}$, 
however, all three models converge, and by the end of the simulation,
the temperatures differ by no more than 10\%. 
We obtain very similar results for free-fall collapse models with 
$n_{\rm i} = 0.03 \: {\rm cm^{-3}}$ and $n_{\rm i} = 30 \: {\rm cm^{-3}}$.

\begin{figure}
\centering
\epsfig{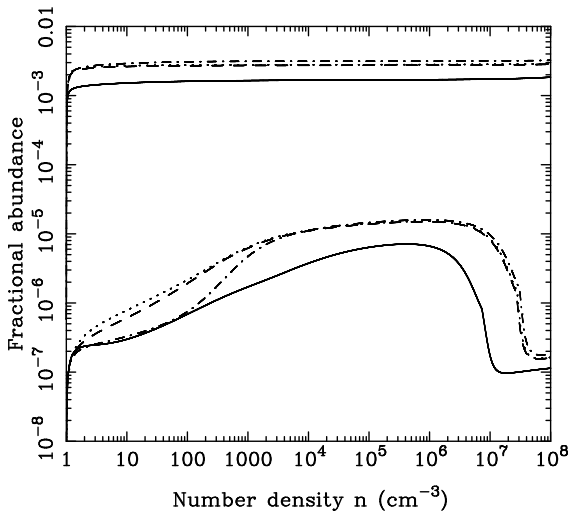}
\caption{Evolution of $\mHt$ and $\hd$ abundances as a function of
gas number density in free-fall collapse models with initial density 
$n_{\rm i} = 1 \: {\rm cm^{-3}}$ and initial redshift $z=20$. We compare
the behaviour for four different treatments of the $\mHt$ cooling:
CF1 (solid line), CF2 (dashed line), CF3 (dash-dotted line) and
CF4 (dotted line). Full details of these treatments are given in the text.
\label{ff_h2}}
\end{figure}

The rapid cooling of the gas at early times in case CF1 is an obvious 
consequence of our inclusion of the cooling arising  from $\mHt$-$\Hp$ 
and $\mHt$-$\me$ collisions, but the relatively slow rate of cooling at 
later times (compared to
the other models) at first seems somewhat counterintuitive: by
adding extra coolants, we have made the gas cool more slowly!
However, this puzzle is easy to solve if we examine the evolution
of the $\mHt$ fraction in these simulations. In Figure~\ref{ff_h2} we show how the
$\mHt$ and $\hd$ abundances evolve for models CF1--CF4; for
clarity, we plot only the $z=20$ case, although the behaviour in the 
$z=10$ case is very similar.

Figure~\ref{ff_h2} shows that significantly less $\mHt$ is produced in
case CF1 than in the other cases, with the difference  amounting to
more than a factor of two at late times. This is a direct result of the 
rapid cooling of the gas at early times in run CF1. Most of the $\mHt$
that forms in all of the runs does so at early times, while the fractional
ionization of the gas is still large. Enhanced cooling during this
period reduces the rate of $\Hm$ formation (owing to the positive
temperature dependence of the reaction coefficient for the formation
of $\Hm$ by radiative association, reaction 1), and also increases
the destruction rate of $\Hm$ by mutual neutralization (since the
range increases for decreasing $T$, regardless of which particular
rate coefficient we adopt). The net result is a reduction in the $\mHt$
formation rate during this critical early period, and hence a reduction
in the $\mHt$ abundance at late times.

Figure~\ref{ff_h2} also demonstrates that this  reduction in the $\mHt$
abundance leads to a corresponding reduction in the HD abundance,
which is a simple consequence of the fact that in most circumstances,
reactions 39 and 41 dominate the production and destruction of HD,
and are in equilibrium, implying that $x_{\hd} \propto x_{\mHt}$.

The differences between runs CF2, CF3 and CF4 are also easy 
to understand. At low densities and low temperatures,
the \citet{GP98} cooling function provides significantly more
cooling per $\mHt$ molecule than the \citet{wf07}, as 
Figure~\ref{H2_cool} demonstrates, and so gas in run CF4
can more easily reach a low temperature than gas in run CF3.
Below a temperature of about 150~K, however, HD cooling begins 
to dominate in both models. As the $\mHt$ abundance does not
differ greatly between the two models, HD cooling is comparably
effective in each, and the thermal evolution of the gas becomes 
insensitive to the choice of $\mHt$ cooling function. The temperature 
curves therefore begin to converge, with this convergence becoming 
complete by the time that the gas  reaches a density $n \simeq 10^{4} 
\: {\rm cm^{-3}}$. At very high densities ($n > 10^{7} \: {\rm cm^{-3}}$), 
the gas once again becomes too warm for HD cooling to dominate, 
as the enhancement of the HD abundance by chemical fractionation 
becomes much less pronounced. However, at these densities, 
$\mHt$ is in LTE, and the only uncertainty in the $\mHt$ cooling rate 
comes from the small uncertainties in the energies of the various 
rotational and  vibrational levels, and in the radiative transition rates. 

Including helium, as in run CF2, increases the $\mHt$ cooling rate,
particularly at low temperatures, and so the gas cools faster. However,
the combination of the \citet{wf07} rate for $\mHt$-$\mH$ cooling with 
the $\mHt$-$\He$ cooling rate presented in \S\ref{h2he_rate}, scaled by the 
appropriate He:H ratio, coincidentally results in a total cooling rate
that is similar to the \citet{GP98} cooling rate: the two differ by no more
than 25\% in the temperature range $210 < T < 1000 \: {\rm K}$
(assuming an ortho-para ratio of 3:1), despite the large disparity in the 
$\mHt$-$\mH$ cooling rates of \citet{GP98} and \citet{wf07} at these
temperatures.

\begin{figure}
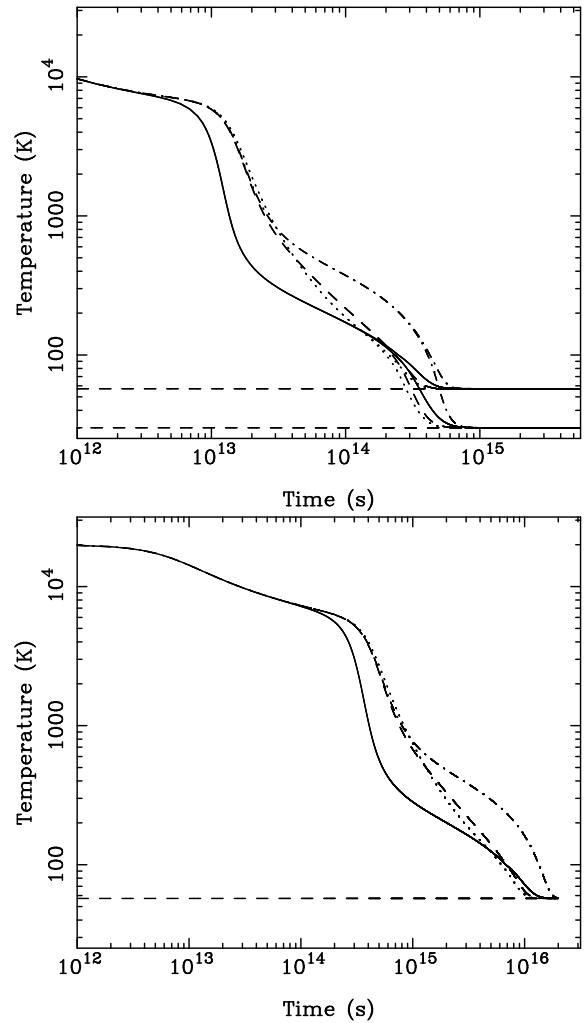

\centering
\epsfig{figure=f8a.eps,width=16pc,angle=-90}
\epsfig{figure=f8b.eps,width=16pc,angle=-90}
\caption{(a) As Figure~\ref{ff_cool}, but showing the temperature
evolution versus time for a set of models in which the evolution of the 
gas is isobaric. The initial density was $n_{\rm i} = 1 \: {\rm cm^{-3}}$, 
and results are plotted for both $z=10$ (lower set of lines) and 
$z=20$ (upper set of lines). Four different treatments of $\mHt$
cooling are compared: CF1 (solid line), CF2 (dashed line), 
CF3 (dash-dotted line) and CF4 (dotted line).  The horizontal
dashed lines indicate the CMB temperature at redshifts $z=10$
and $z=20$. For reference, the 
Hubble time at $z=10$ is $t_{\rm H} \simeq 2.3 \times 10^{16} \: {\rm s}$ and
at $z=20$ is $t_{\rm H} \simeq 9.0 \times 10^{15} \: {\rm s}$, where we 
have adopted the standard WMAP3 cosmological parameters 
\citep{wmap3}.
(b) As (a), but for isobaric models with $n_{\rm i} = 0.03 \: {\rm cm^{-3}}$;
note that only the $z=20$ case is plotted. \label{iso_cool}}
\end{figure}

The isobaric evolution models tell a similar story. Figure~\ref{iso_cool}a,
shows how the gas temperature evolves with time in two representative
models with initial density $n_{\rm i} = 1 \: {\rm cm^{-3}}$ and redshifts
$z=10$ (lower curves) and $z=20$ (upper curves). We again find that
the choice of $\mHt$ cooling function affects the temperature evolution.
Models CF2 and CF4 again barely differ, while model CF3 differs from
them significantly only at low temperatures ($T < 700 \: {\rm K}$). At
high temperatures ($T \gg 700 \: {\rm K}$), cooling in all three of these
models is dominated the vibrational excitation of $\mHt$ by hydrogen
atoms, and the vibrational rates differ little between the \citet{GP98}
and \citet{wf07} treatments. On the other hand, if we include the effects
of $\Hp$ and $\me$ collisional excitation of $\mHt$, as in model CF1, 
we see a more substantial difference in the temperature evolution of
the gas, at all temperatures $T < 7000 \: {\rm K}$. Because of the high
initial ionization, $\mHt$-$\Hp$ collisions dominate, and the gas cools
much faster than in the other models. We find similar results in 
our $n_{\rm i} = 0.03 \: {\rm cm^{-3}}$ and $n_{\rm i} = 30 \: {\rm cm^{-3}}$ 
models. 

The importance of the difference in cooling time is unclear, but likely 
depends upon the other relevant timescales in the problem. For instance,
in the present example, the time required for the gas to cool to $T_{\rm CMB}$ 
is much less than the Hubble time $t_{\rm H}$ regardless of which treatment 
of $\mHt$ cooling is used. On the other hand, in our $n_{\rm i} = 0.03 \: 
{\rm cm^{-3}}$, $z=20$ model, illustrated in Figure~\ref{iso_cool}b, we 
find that $t_{\rm cool} \simeq t_{\rm H}$ for CF1, CF2 and CF4, but is 
approximately twice as long as the Hubble time if we use treatment CF3.

We close this section by noting that despite the differences in the
temperature evolution brought about by a change in $\mHt$
cooling function, in every case we have examined the gas remains
able to cool below the temperature reachable by $\mHt$ cooling
alone. In other words, HD cooling is important in every case 
considered here, and the gas always reaches the regime in which
HD cooling dominates. However, in the free-fall models, the 
minimum temperature reached by the gas is always higher than
the temperature floor set by the CMB, and varies depending
on whether or not we include the effects of $\mHt$-$\Hp$ and
and $\mHt$-$\me$ collisions when computing the $\mHt$
cooling rate. If, as has been hypothesized by some authors 
\citep[see e.g.][]{jb06}, gravitational fragmentation of the gas 
occurs only once the gas reaches its minimum temperature, 
then simulations that do not include these processes will
produce fragments that are too small by roughly  a factor of
two. On the other hand, if the outcome of the fragmentation 
process is determined in part by the gas dynamics at earlier 
times, then the inclusion of these processes could conceivably
enhance fragmentation, owing to the reduction in the cooling
time of the gas at early times, and the much flatter temperature
dependence of the $\mHt$-$\Hp$ cooling rate compared to
the $\mHt$-$\mH$ cooling rate. Our very simple dynamical models 
do not allow us to explore these issues in any greater detail, 
but will hopefully motivate further work on the subject.

\subsection{Uncertainties in the reaction rate coefficients}
\label{res-chem}
In \S\ref{chem-model}, we discussed the large uncertainties that exist in
some of the rate coefficients for reactions included in our
chemical model. The most uncertain rates in our model are
the destruction of $\Hm$ by associative detachment with $\mH$ 
(reaction 2; see \S\ref{admn-rates}), the destruction of $\Hm$ by mutual 
neutralization with $\Hp$ (reaction 5; again see \S\ref{admn-rates}),
the destruction of $\mHt$ by charge transfer with $\Hp$ (reaction 7; see \S\ref{h2ct-rates}),
and the three-body formation of $\mHt$ (reactions 30 and 31; see \S\ref{h2tb-rates}).  
In the following sections, we examine the individual effects of each of
these uncertainties, before concluding by placing limits on the combined
effect of all four uncertainties. 

Although we have examined the effects of these uncertainties for all of the
combinations of $n_{\rm i}$ and $z$ considered previously in this paper,
for simplicity (and for clarity in the figures) we restrict our discussion here 
to one particular case: $n_{\rm i} = 1 \: {\rm cm^{-3}}$ and $z=20$. Unless
otherwise noted, we find very similar results for all of the other combinations
of redshift and density that we have studied.

\subsubsection{Associative detachment and mutual neutralization}
\begin{figure}
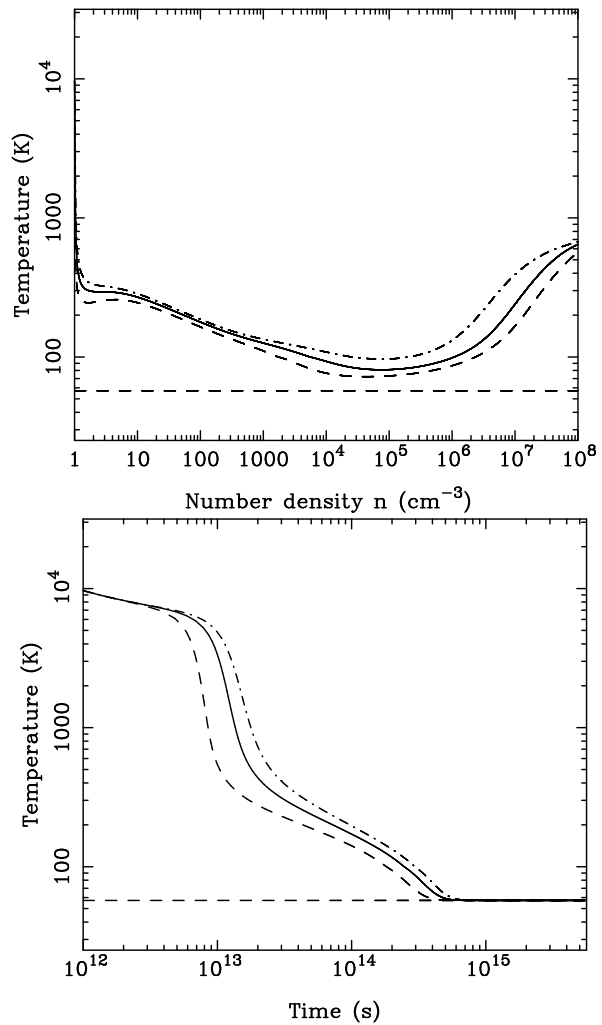

\centering
\epsfig{figure=f9a.eps,width=16pc,angle=-90} 
\epsfig{figure=f9b.eps,width=16pc,angle=-90}  
\caption{(a) Temperature evolution as a function of gas density for 
a free-fall collapse model with $n_{\rm i} = 1\: {\rm cm^{-3}}$ and
$z=20$. Three different values are used for the $\Hm$ associative
detachment rate coefficient ($k_{2}$): our default value, taken from
\citet{sff67} (solid line); and `maximal' and `minimal' values
(dashed and dash-dotted lines, respectively)  taken from \citet{gsj06}.
The horizontal dashed line indicates the CMB temperature at
$z=20$.
(b) As (a), but showing the time evolution of temperature in an 
isobaric model with the same initial conditions. \label{adfig}}
\end{figure}

\begin{figure}
\centering
\epsfig{figure=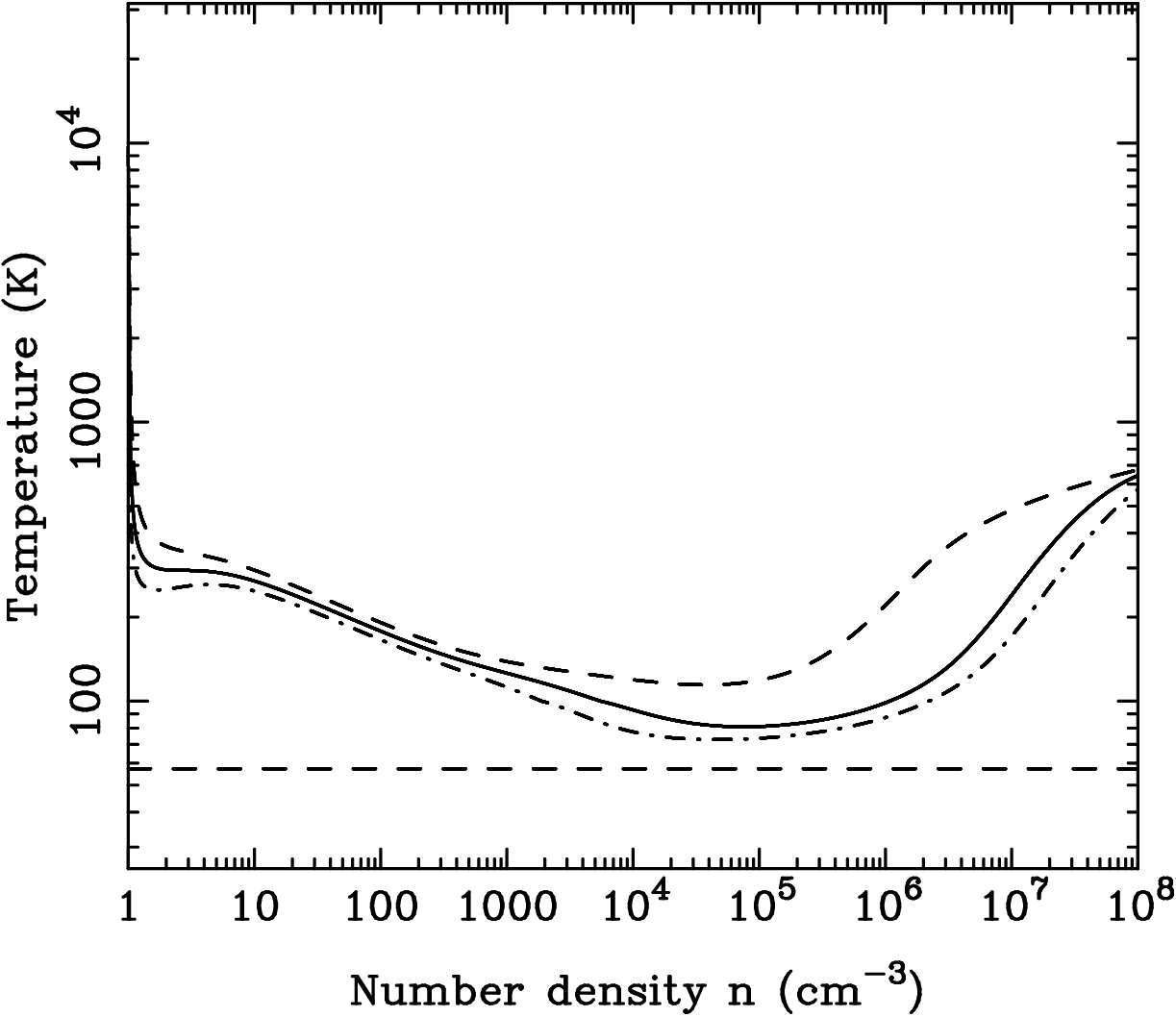,width=16pc,angle=-90}  
\epsfig{figure=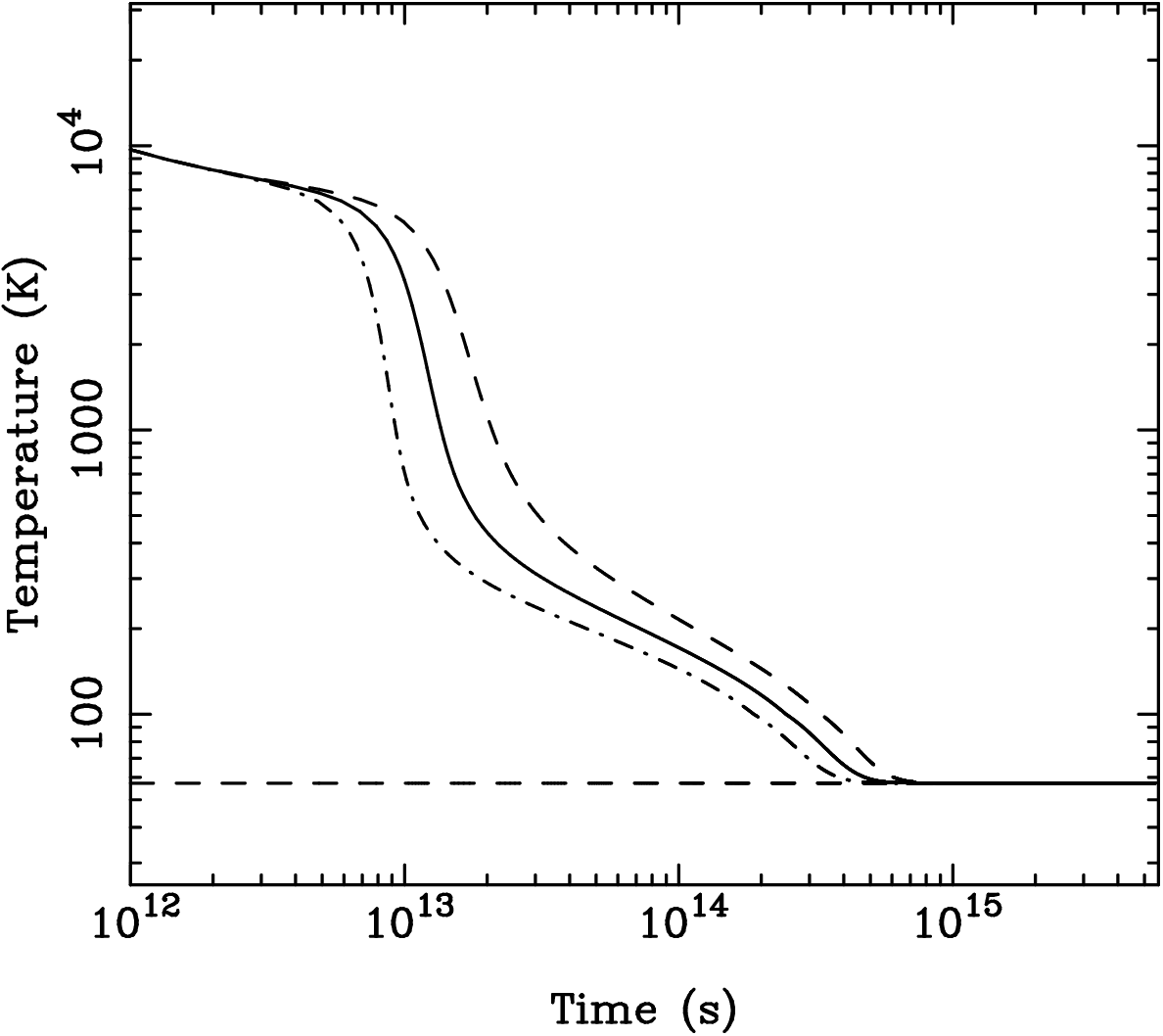,width=16pc,angle=-90}  
\caption{(a) Temperature evolution as a function of gas density for 
a free-fall collapse model with $n_{\rm i} = 1\: {\rm cm^{-3}}$ and
$z=20$. Three different values are used for the $\Hm + \Hp$ 
mutual neutralization rate coefficient ($k_{5}$): our default rate
(solid line; CDG99), taken from \citet{cdg99}, along with a large rate
(dashed line; MOS70), taken from \citet{MOS70}, and a smaller rate
(dash-dotted line; DL87), taken from \citet{dl87}. 
The horizontal dashed line indicates the CMB temperature at
$z=20$.
(b) As (a), but showing the time evolution of temperature in an 
isobaric model with the same initial conditions.
\label{mnfig}}
\end{figure}

In gas cooling and recombining from an initially ionized state, the
amount of $\mHt$ formed is sensitive to the ratio between the 
destruction rate of $\Hm$ by associative detachment with $\mH$
(reaction 2) and by mutual neutralization with $\Hp$ (reaction 5).
An increase in $k_{2}$ or a decrease in $k_{5}$ leads to associative
detachment becoming the dominant destruction process at earlier
times, when the fractional ionization of the gas is larger, and hence
leads to a larger final $\mHt$ fraction; conversely, a decrease in
$k_{2}$ or an increase in $k_{5}$ means that mutual neutralization
dominates for a longer period, and hence the final $\mHt$ fraction
is smaller. As \citet{gsj06} have already shown, in the absence of
a substantial ultraviolet background, the effect on the final $\mHt$
fraction is not as large as might be feared: an order of magnitude
change in $k_{2}$ or $k_{5}$ alters the final $\mHt$ fraction by
no more than a factor of a few. Nevertheless, this is enough to 
alter the temperature evolution of the gas by an appreciable 
amount, as we can see from Figures~\ref{adfig} and \ref{mnfig}.

Figure~\ref{adfig}a shows the effect on a representative free-fall 
collapse model of varying the associative detachment rate while 
keeping the mutual neutralization rate fixed. We plot results from
a model using our default value for $k_{2}$, taken from \citet{sff67},
and from models using  `maximal' and `minimal' values for $k_{2}$
taken from \citet{gsj06}. The uncertainty 
in $k_{2}$ introduces an uncertainty into the temperature 
evolution that persists throughout the simulation.  Gas in simulations with 
a high value for $k_{2}$ (and hence higher $\mHt$ fractions) has 
a systematically lower temperature than the gas in simulations with a low 
value for  $k_{2}$. The difference between the simulations is
particularly pronounced for densities in the range $10^{6} < n < 10^{7} \:
{\rm cm^{-3}}$. The relatively rapid increase in the gas temperature 
at these densities occurs because HD reaches local thermodynamic
equilibrium, and hence can no longer cool the gas so effectively.
The causes the gas to begin heating up, which in turn reduces
the HD abundance (as fractionation becomes less effective),
causing the gas to warm further. The rate at which this process
occurs depends upon the initial HD abundance, and hence on
the $\mHt$ abundance; reheating occurs more slowly when the
$\mHt$ abundance is large. The gas temperature in this density
regime can therefore differ by a factor of two or more, depending
on which value is chosen for  $k_{2}$.

In Figure~\ref{adfig}b, we show how the same variation in
$k_{2}$ affects the temperature evolution in a representative
isobaric model. In this case, the reduction in the $\mHt$
fraction resulting from a decrease in $k_{2}$ systematically
delays cooling relative to our reference calculation. Similarly,
an increase in $k_{2}$ accelerates cooling. In the present
context, what is perhaps most interesting is the time taken to
reach the temperature floor set by the CMB. This occurs
after $t \simeq 4 \times 10^{14} \: {\rm s}$ in the model with
the largest value of $k_{2}$ and after $t \simeq 6 \times
10^{14} \: {\rm s}$ in the model with the smallest value
of $k_{2}$.  We find a similar degree of  uncertainty in the 
cooling times in our other isobaric models.

In Figure~\ref{mnfig}, we examine the effect of varying the
mutual neutralization rate while keeping the associative
detachment rate fixed. We plot results from models performed
using mutual neutralization rates from \citet{cdg99} -- our
default -- as well as from \citet{MOS70} and \citet{dl87}.
Figure~\ref{mnfig} shows that varying the mutual 
neutralization rate has very similar effects to varying the 
associative detachment rate, except that the sense of the
effect is reversed: a decrease in $k_{5}$ has a similar
effect to an increase in $k_{2}$ and vice versa. The size of 
the uncertainty introduced into the temperature evolution 
of the gas is comparable at low densities, and somewhat
larger at high densities, where the gas temperatures differ
by as much as a factor of four.

Nevertheless, although both rate coefficient uncertainties clearly
affect the cooling of the gas, in neither case do they substantially
change the outcome of the simulations. The gas still cools to 
temperatures low enough for chemical fractionation to significantly
enhance HD, and so in each case HD cooling becomes dominant,
further cooling the gas. In our free-fall collapse models, the minimum
temperature reached by the gas does depend on the values of $k_{2}$
and $k_{5}$, and this may affect the characteristic fragment mass
scale, although we would still expect any fragments to be smaller than
in the case where only $\mHt$ cooling is effective. In our isobaric
models, the same minimum temperature is reached in every case,
but the time taken to arrive there differs by up to a factor of two.
The mass accretion dependent collapse times found by \cite{yahs03} 
suggest that such a factor of two uncertainty can be relevant,
but further investigation requires a proper three-dimensional
hydrodynamical treatment; our highly simplified dynamical models 
can take us no further.

\subsubsection{$H_{2}$ charge transfer}
\begin{figure}
\centering
\epsfig{figure=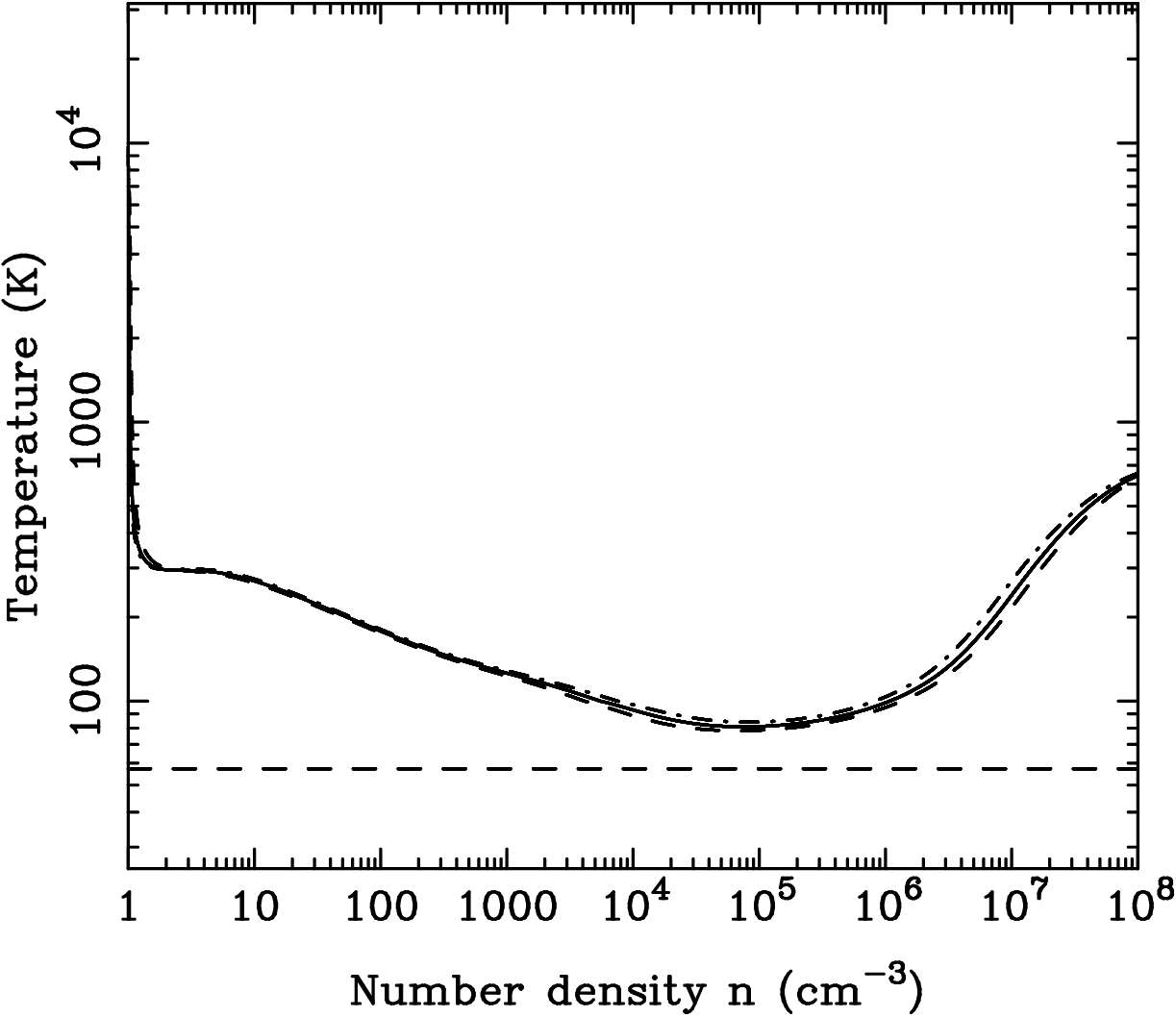,width=16pc,angle=-90} 
\epsfig{figure=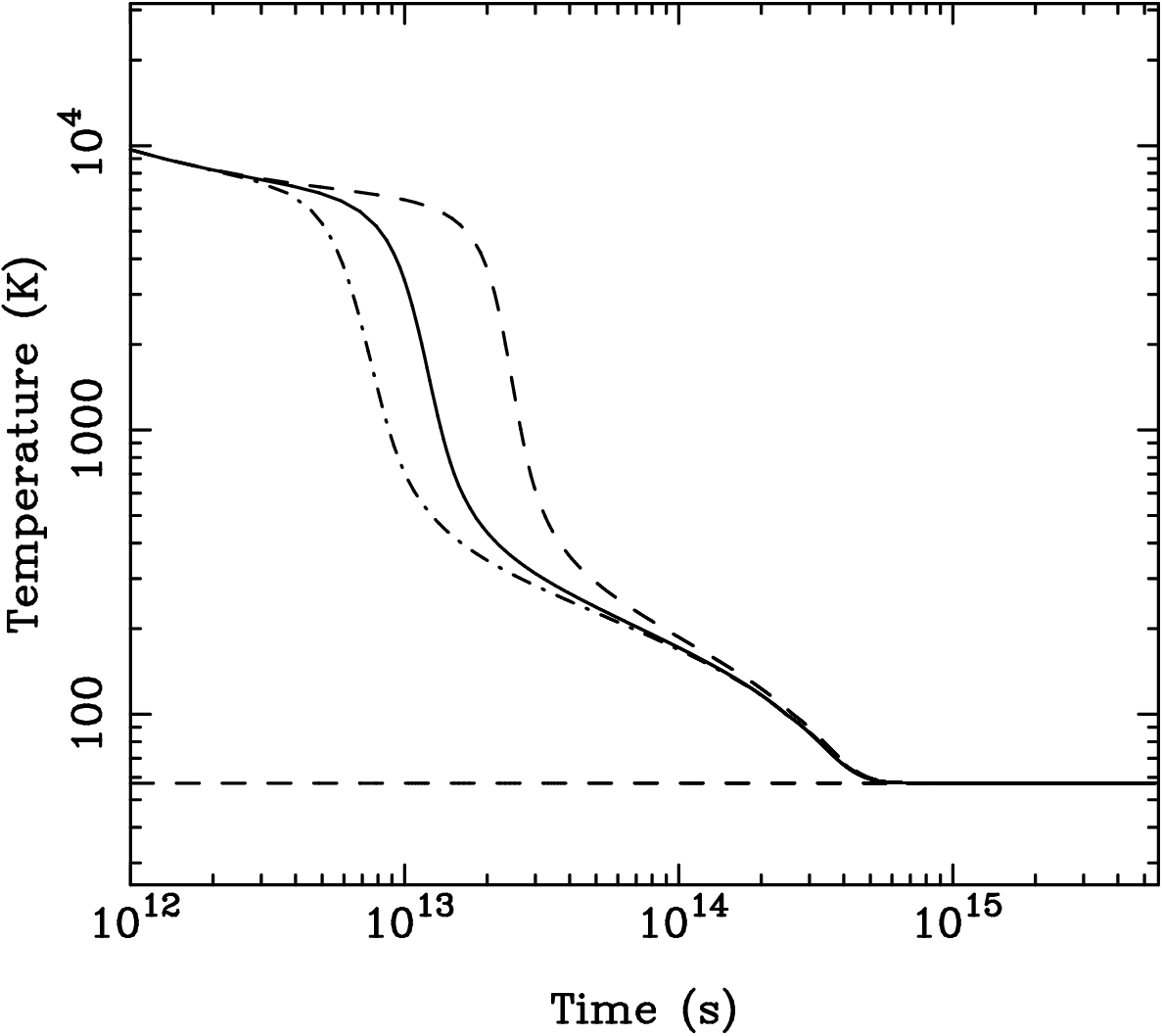,width=16pc,angle=-90} 
\caption{(a) Temperature evolution as a function of gas density for 
a free-fall collapse model with $n_{\rm i} = 1\: {\rm cm^{-3}}$ and
$z=20$. Three different values are used for the rate coefficient 
for $\mHt$ destruction by charge transfer with $\Hp$ (reaction 7).
The rate used in our reference model (solid line) is taken from
\citet{SAV04}, but we also show the effects of using rates from
\citet{sk87} (dashed line) and \citet{ABE97} (dash-dotted line).
The horizontal dashed line indicates the CMB temperature at
$z=20$.
(b) As (a), but showing the time evolution of temperature in an 
isobaric model with the same initial conditions. 
\label{ctfig}}
\end{figure}

In Figure~\ref{ctfig}, we examine the impact of varying the rate
coefficient for $\mHt$ destruction by charge transfer ($k_{7}$)
in the context of representative free-fall collapse and isobaric
evolution models. In the free-fall model, the effect of increasing
$k_{7}$ is to enable the gas to cool to lower temperatures. At
first sight, this seems counterintuitive: by destroying $\mHt$,
we make the gas colder. However, the key is that charge transfer
is only an effective destruction mechanism at high temperatures.
By increasing the charge transfer rate, we delay the onset of rapid
$\mHt$ cooling, and so when the gas does become able to cool
rapidly, the fractional ionization is lower, and the cooling time is
longer. Consequently, the gas remains warm for a longer period,
forms more $\Hm$ (owing to the temperature dependence of 
reaction 1), and hence forms more $\mHt$. Nevertheless, the
effect is relatively small: the uncertainty in the $\mHt$ fraction 
once the gas has cooled is no more than 5\%, and the uncertainty
in the minimum temperature is no more than 10\%.

In the isobaric model, we again see that the effect of increasing
$k_{7}$ is to delay the onset of efficient $\mHt$ cooling. However,
once the gas begins cooling, the temperature evolution becomes
convergent and the final outcome of the simulations is insensitive
to the value of $k_{7}$.

\subsubsection{Three-body $H_{2}$ formation}
\begin{figure}
\centering
\epsfig{figure=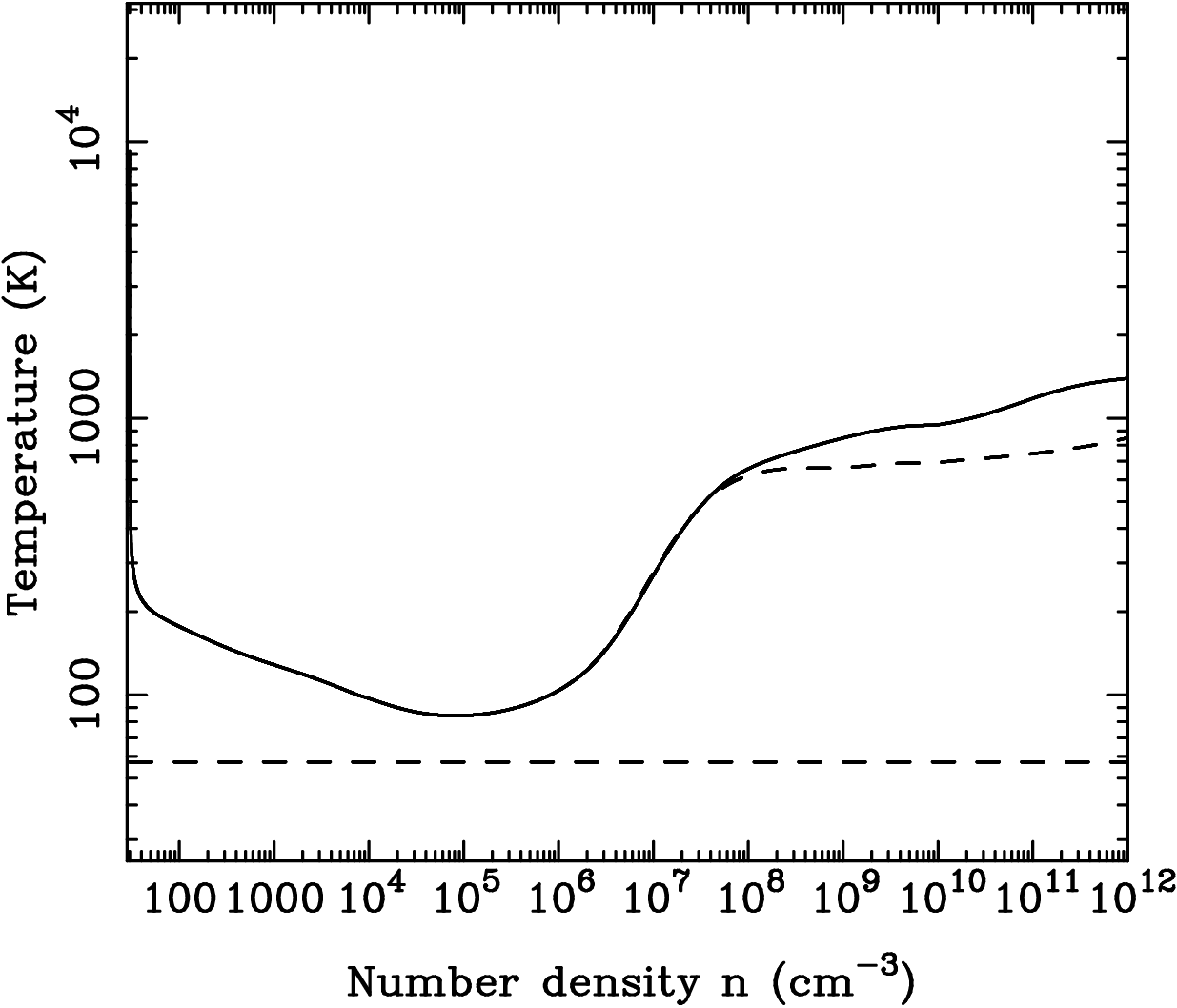,width=16pc,angle=-90}  
\epsfig{figure=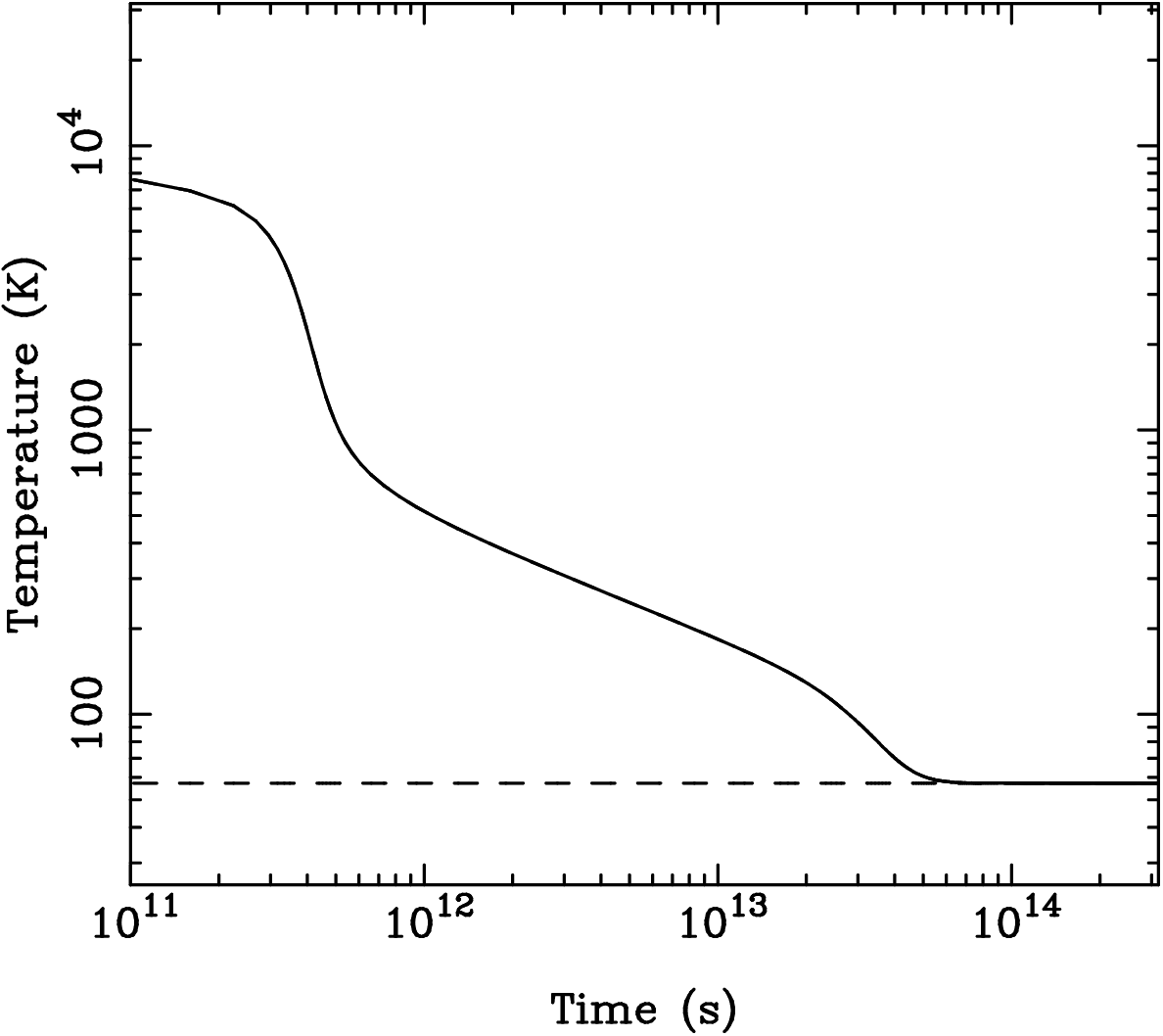,width=16pc,angle=-90}  
\caption{(a) Temperature evolution as a function of gas density in
two simulations with $n_{\rm i} = 30 \: {\rm cm^{-3}}$ and $z=20$. Two different
values are used for the rate coefficient for three-body $\mHt$ formation
($k_{30}$): our default value (solid line), taken from \citet{abn02}, 
and a much larger value (dashed line) recently computed by \citet{fh07}.
The horizontal dashed line indicates the CMB temperature at
$z=20$.
(b) As (a), but showing the time evolution of temperature in an 
isobaric model with the same initial conditions. Note that in this
plot, the results of the two simulations are indistinguishable.
\label{h2tb_fig}}
\end{figure}

Since we have assumed, following \citet{pss83}, that the rate 
coefficients for reactions 30 and 31 are related by $k_{31} = 
k_{30} / 8$, we can explore the effects of the uncertainty in
the three-body rates simply by varying $k_{30}$. In Figure~\ref{h2tb_fig}, 
we examine the effect of the uncertainty in reaction 30 in two 
representative models: one free-fall collapse model and one isobaric 
model. 

In the free-fall collapse model, the effect of the uncertainty is apparent
only for densities $n > 5 \times 10^{6} \: {\rm cm^{-3}}$. Between this
density and $n = 10^{8} \: {\rm cm^{-3}}$, the gas temperature increases
slightly faster in the simulation that uses the larger \citet{fh07} rate than in the 
simulation that uses the smaller \citet{abn02} rate, owing to the greater three-body 
$\mHt$ formation heating rate in the former case. At $n > 10^{8} \: {\rm cm^{-3}}$,
however, the greater heating rate in the \citeauthor{fh07} run is more than 
counterbalanced by the greater cooling provided by the larger abundance 
of $\mHt$, and so the gas temperature increases at a slower rate than 
in the \citeauthor{abn02} run. This difference persists until we terminate
the simulation at $n = 10^{12} \: {\rm cm^{-3}}$, and the final temperatures
differ by about 65\%.

In the isobaric model, there are no obvious differences between the
two simulations. This is to be expected, as the gas density in these 
simulations never 
exceeds a few thousand particles per cubic centimeter, and so the
three-body $\mHt$ formation rate remains extremely small throughout
both simulations.

In neither case does the uncertainty affect the cooling of the gas; in
the free-fall model, it merely affects how quickly the gas subsequently 
re-heats. Therefore, whatever its impact on later stages of the star
formation process,  from the point of view of understanding the role and 
effectiveness of HD cooling, this particular source 
of uncertainty is clearly unimportant. 

\subsubsection{Combining the uncertainties}
\begin{figure}
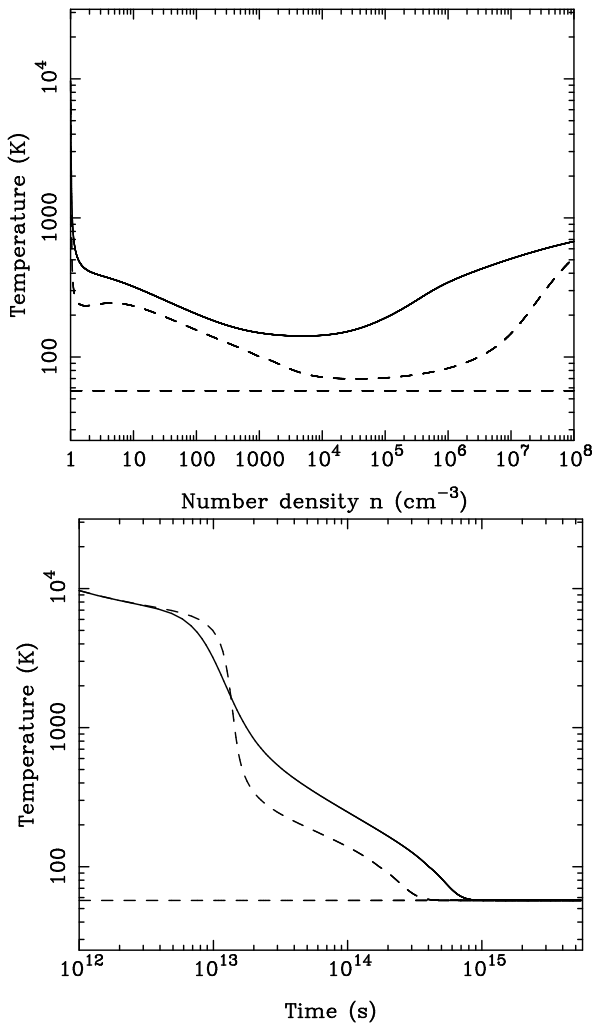

\centering
\epsfig{figure=f13a.eps,width=16pc,angle=-90} 
\epsfig{figure=f13b.eps,width=16pc,angle=-90} 
\caption{(a) Temperature evolution as a function of gas density
in two simulations with $n_{\rm i} = 1 \: {\rm cm^{-3}}$ and $z=20$, with
values for the uncertain rate coefficients selected so as to
minimize (solid line) or maximize (dashed line) the amount of
cooling from $\mHt$.
The horizontal dashed line indicates the CMB temperature at
$z=20$. 
(b) As (a), but showing the time evolution of temperature in an 
isobaric model with the same initial conditions.
\label{minmax}}
\end{figure}

We close this discussion by examining two limiting cases, illustrated in
Figure~\ref{minmax},  where we have selected the values of the various 
uncertain rate coefficients in order to maximize and to minimize the 
degree of  cooling. In our `maximal' model, we used a value of 
$5.0 \times 10^{-9} \: {\rm cm^{3}} \: {\rm s^{-1}}$ for the $\Hm$ associative
detachment rate coefficient \citep{gsj06}, and used values for the $\Hm$ mutual
neutralization, $\mHt$ charge transfer and $\mHt$ three-body formation
rate coefficients from \citet{dl87}, \citet{ABE97} and \citet{abn02} respectively. 
In our `minimal' model, we used a value of $0.65 \times 10^{-9} \: {\rm cm^{3}} 
\: {\rm s^{-1}}$ for the $\Hm$ associative detachment rate coefficient 
\citep[again from][]{gsj06}, along with values for the other three rate 
coefficients taken from \citet{MOS70}, \citet{sk87} and \citet{fh07} respectively.
Note that in both models, the \citet{wf07} $\mHt$ cooling rate is used
and the effects of cooling from $\mHt$-$\mHt$, $\mHt$-$\He$, $\mHt$-$\Hp$
and $\mHt$-$\me$ collisions are included; i.e.\ we do not couple the
chemical rate uncertainties to the cooling rate uncertainties.

In Figure~\ref{minmax}a, we examine the effects of the `minimal' and 
`maximal' models in the context of a representative free-fall collapse
calculation with an initial redshift $z=20$. The temperature evolution 
in the two models differs significantly at all densities, particularly in
the interval $10^{4} < n < 10^{8} \: {\rm cm^{-3}}$. The minimum 
temperature reached by the gas in the `maximal' model is 69.8~K,
and in the `minimal' model is 141.5~K, a factor of two difference
(corresponding to a difference in the Jeans mass at this point of
almost a factor of three). It is clear that in the `minimal', HD cooling
is of limited importance: although some HD cooling does occur, it
never becomes completely dominant, and the gas temperature
remains well above the floor set by the CMB. If the rates adopted
in this model do prove to be the most accurate ones, then this
would imply that HD cooling in collapsing gas may be significantly
less effective than previously thought.

Finally, in Figure~\ref{minmax}b, we examine the effects of the 
`minimal' and `maximal' models in the context of an isobaric 
evolution model with an initial redshift $z=20$. Again, we see 
that the rate coefficient uncertainties significantly affect the cooling 
time of the gas, lengthening it in this case by a factor of three.
However, they do not change the final outcome of the simulation:
the gas still cools down to 
$T \sim T_{\rm CMB}$, and even for the `minimal' model, the time
required to reach $T_{\rm CMB}$ when $n_{\rm i} = 1 \: {\rm cm^{-3}}$
is significantly less than a Hubble time.

\section{Summary}
\label{concl}
We have detailed the effect of several possible sources of uncertainty
on the thermal evolution of primordial gas cooling from an initially
hot and ionized state. These considered potential sources of
uncertainty are the sensitivity of the low temperature $\mHt$ cooling
rate to the ratio of ortho-$\mHt$ to para-$\mHt$ (which may differ
significantly at low temperatures from the 3:1 ratio that is usually
assumed), the continuing uncertainty in the form of the low
temperature cooling rate for $\mHt$ excited by collisions with atomic
hydrogen, the neglect by most previous authors of the contributions
made to the $\mHt$ cooling rate by collisions of $\mHt$ with $\He$,
$\Hp$ and $\me$, and the large uncertainties that exist in the rates
of several of the reactions responsible for determining the $\mHt$
fraction in the gas.

We find that the first of these sources of uncertainty is unimportant.
The standard assumption of a 3:1 ortho-para ratio is reasonably
accurate at temperatures where $\mHt$ cooling is effective. It only
becomes significantly inaccurate for temperatures $T \simless 100 \:
{\rm K}$, but at these temperatures $\mHt$ cooling is unimportant in
comparison to HD cooling, and so this inaccuracy has no effect on the
thermal evolution of the gas. However, the presented rates and
formalism should prove useful when detailed complete emission models
from cooling primordial gas are to be constructed.

The second source of uncertainty -- i.e.\ whether one uses the
popular \citet{GP98} parameterization to represent the cooling
rate due to $\mHt$-$\mH$ collisions, or the newer rates of
\citet{wf07} -- can give the impression of being important if
one assumes that this process dominates the total $\mHt$
cooling rate. At low temperatures ($T < 1000 \: {\rm K}$),
the \citet{wf07} cooling function provides significantly less
cooling than the \citet{GP98} cooling function, and the use
of the former in place of the latter lengthens the cooling time
of the gas by about a factor of two.

However, this simple comparison overstates the effect of
this uncertainty. The problem stems from the assumption
that $\mHt$-$\mH$ collisions dominate. In practice, if one
adopts the \citet{wf07} cooling rate, then at low temperatures,
$\mHt$-$\He$ collisions are more important, despite the low
abundance of He relative to H. If one includes the effects
of these collisions in the calculations,
then the sensitivity of the outcome to the uncertainty
in the $\mHt$-$\mH$ rate becomes considerably less.

A more important source of error in previous investigations
of HD formation in relic \hii regions and other similar
environments is the neglect of the effects of cooling due
to collisions between $\mHt$ and protons and electrons.
At early times during the cooling of the gas, the fractional
ionization remains large, and $\mHt$-$\Hp$ collisions
dominate. When their effects are included, the cooling
time of the gas is significantly decreased. However, a
side effect of this rapid cooling is that less $\mHt$ forms,
owing to the temperature dependence of the $\mHt$
formation rate. Because of this, less HD is formed,
and so HD cooling is less effective. Therefore, the gas
remains significantly warmer at later times than it would
in models that did not include cooling from  $\mHt$-$\Hp$
collisions, although it remains cooler than it would be
if HD cooling were not included. The increase in the
minimum temperature in our free-fall models corresponds
to an increase in the minimum Jeans mass of a factor of
two, suggesting that previous studies of population III
star formation in formerly-ionized regions may have
underestimated the characteristic mass of the stars that
form. Moreover, previous work by \citet{yahs03}
has shown that variations in the cooling time of the gas at
low densities can have a pronounced effect on the supply
of cold gas available for population III star formation, and
on the timing of the collapse. Furthermore, \citet{oshn07}
have shown that the amount of $\mHt$ formed at low densities,
which is sensitive to the thermal evolution of the gas, can affect
the eventual accretion rate of gas onto the protostar,
and hence may also affect its final mass. We therefore
anticipate that the inclusion of the effects of $\mHt$-$\Hp$
cooling may lead to clear differences in the outcome of such
three-dimensional studies.

Our investigation into the effects of the chemical rate coefficient
uncertainties has shown that the large uncertainties in the
associative detachment and mutual neutralization rates also have a
significant impact on the thermal evolution of the gas. Although HD
becomes the dominant coolant in all of the models considered,
variations in the associative detachment and mutual neutralization
rates alter the minimum temperature reached by the gas in our
free-fall collapse simulations, and have a particularly pronounced
effect on the temperature evolution at the end of the period during
which HD cooling dominates. It is therefore quite plausible that these
uncertainties may also modify the outcome of multi-dimensional
hydrodynamical simulations.

On the other hand, our results display only a small sensitivity
to the uncertainty in the $\mHt$ charge transfer reaction, other
than at very early times in the isobaric runs, due to
the fact that most of the $\mHt$ that forms in the gas
does so at temperatures where this reaction is ineffective.
Further investigation of the effects of this uncertainty
may be of interest, but is clearly not a high priority.

Finally, our results demonstrate that the evolution of gas in the
HD-cooled regime is insensitive to the large uncertainty that exists
in the three-body $\mHt$ formation rate, as three-body processes are
unimportant at the densities at which HD dominates. However, this
uncertainty does become important at densities $n > 10^{8} \: {\rm
   cm^{-3}}$. Its effect on the hydrodynamics of the gas at these
densities remains uncertain.

In summary, we discussed a number of hitherto neglected physical
processes. The most crucial finding is that the molecular hydrogen 
cooling initiated by collisions with electrons, protons and neutral 
helium cannot be neglected in general, and should be included in 
future studies of the dynamics of primordial gas.

\section*{Acknowledgments}
The authors would like to thank D.~Galli for discussions concerning $\mHtp$
cooling, and for kindly providing to us the fit given in Eq.~\ref{galli_fit}.
They would also like to thank A.~Dalgarno, N.~Yoshida, D.~Whalen and the anonymous
referee for their comments on an earlier draft of this article. In particular, the referee 
deserves thanks for pointing out to us the possible importance of cooling from 
collisions between molecular hydrogen and helium. SCOG acknowledges 
useful discussions with D.~Savin and P.~Stancil regarding the
rate of reactions involving deuterium. Financial support was provided in part by the
US National Science Foundation under Grant No.\ PHY05-51164, and by
NSF CAREER award AST-0239709. This project was
initiated while the authors were participants in the workshop ``Star Formation
Through Cosmic Time'' at the Kavli Institute for Theoretical Physics, and we thank
the staff and scholars of KITP for their hospitality.

\appendix
\section{Chemical network}
In Table~\ref{chem_gas} we list the chemical reactions included in our model 
of primordial gas, along with the rate coefficients adopted and the references 
from which these rates were taken. For further details on some of the reactions, 
see also \S\ref{chem-model}.

\begin{table*}
\begin{minipage}{126mm}
\caption{List of reactions included in our chemical model
\label{chem_gas}}
\begin{tabular}{llllc}
\hline 
No.\  & Reaction & Rate coefficient $({\rm cm}^{3} \: {\rm s}^{-1})$ & & Ref.\ \\
\hline
1  &  $\mH + \me  \rightarrow  \Hm + \gamma $ &
$k_{1} = {\rm dex}[-17.845 + 0.762 \log{T}$ & $T \le 6000 \: {\rm K}$ & 1 \\
& & $\phantom{k_{1} = {\rm dex}[} \mbox{}+ 0.1523 (\log{T})^{2}$ & & \\
& & $\phantom{k_{1} = {\rm dex}[} \mbox{}- 0.03274 (\log{T})^{3}] $ & & \\
& & $ \phantom{k_{1}} = {\rm dex}[-16.4199 + 0.1998 (\log{T})^{2}$ & $T > 6000 \: {\rm K}$ & \\ 
& & $ \phantom{k_{1} = {\rm dex}[} \mbox{}-5.447  \times 10^{-3}  (\log{T})^{4}$ & & \\ 
& & $ \phantom{k_{1} = {\rm dex}[} \mbox{}+ 4.0415 \times 10^{-5} (\log{T})^{6}]$  & & \\
2 & $\Hm  + \mH  \rightarrow \mHt + \me$ & See text & & --- \\
3  &  $\mH + \Hp  \rightarrow  \mHtp + \gamma $ &
$k_{3} = {\rm dex}[-19.38 - 1.523 \log{T} $ & & 2 \\
& & $\phantom{k_{3}=} \mbox{} + 1.118 (\log{T})^{2}  - 0.1269 (\log{T})^{3}]$ & & \\ 
4 & $\mH + \mHtp \rightarrow \mHt + \Hp$ & $k_{4} = 6.4 \times 10^{-10}$ & & 3 \\
5 & $\Hm  + \Hp  \rightarrow \mH + \mH$ & See text  & & --- \\
6  &  $\mHtp + \me  \rightarrow  \mH + \mH $ &
$k_{6} = 1.0 \times 10^{-8}$  & $T \le 617 \: {\rm K}$ & 4 \\
& & $\phantom{k_{6}} = 1.32 \times 10^{-6} T^{-0.76}$ & $T > 617 \: {\rm K}$ & \\
7  &  $\mHt + \Hp  \rightarrow  \mHtp + \mH $ &
$k_{7} = [- 3.3232183 \times 10^{-7}$ & & 5 \\
& & $\phantom{k_{7} =}  \mbox{} + 3.3735382 \times 10^{-7}  \log{T}$  & & \\
& & $\phantom{k_{7} =}  \mbox{} - 1.4491368 \times 10^{-7}  (\log{T})^2$ & & \\
& & $\phantom{k_{7} =}  \mbox{} + 3.4172805 \times 10^{-8}  (\log{T})^3$ & & \\
& & $\phantom{k_{7} =}  \mbox{} - 4.7813720 \times 10^{-9}  (\log{T})^4$ & & \\
& & $\phantom{k_{7} =}  \mbox{} + 3.9731542 \times 10^{-10} (\log{T})^5$ & & \\
& & $\phantom{k_{7} =}  \mbox{}  - 1.8171411 \times 10^{-11}  (\log{T})^6$ & & \\
& & $\phantom{k_{7} =}  \mbox{}  + 3.5311932 \times 10^{-13} (\log{T})^7 ]$ & & \\
& & $\phantom{k_{7} =} \mbox{} \times \exp \left(\frac{-21237.15}{T} \right)$ & & \\
8 & $\mHt + \me  \rightarrow  \mH + \mH +  \me$ &
$k_{8} = 4.49 \times 10^{-9} T^{0.11} \expf{-}{101858}{T}$ & $v=0$ & 6 \\
& & $\phantom{k_{8}} = 1.91 \times 10^{-9} T^{0.136} \expf{-}{53407.1}{T}$ & LTE & 6 \\
9 & $\mHt + \mH  \rightarrow  \mH + \mH + \mH$  & 
$k_{9} = 6.67 \times 10^{-12} T^{0.5} \exp \left[-(1+ \frac{63593}{T}) \right]$ & v=0 & 7 \\
& & $\phantom{k_{9}} = 3.52 \times 10^{-9} \expf{-}{43900}{T}$ & LTE & 8 \\
10 & $\mHt + \mHt \rightarrow  \mHt + \mH + \mH$ & 
$k_{10} = \frac{5.996 \times 10^{-30} T^{4.1881}}{(1.0 + 6.761 \times 10^{-6} T)^{5.6881}}
\exp \left(-\frac{54657.4}{T} \right)$ & $v=0$ & 9 \\
& & $\phantom{k_{10}} = 1.3 \times 10^{-9} \expf{-}{53300}{T}$ & LTE & 10 \\
11 & $\mHt + \He \rightarrow \mH + \mH + \He$ & 
$k_{11} = {\rm dex} \left[ -27.029 + 3.801 \log{T} - \frac{29487}{T} \right]$ & $v=0$ & 11 \\
& & $\phantom{k_{11}} =  {\rm dex} \left[ -2.729 -1.75 \log{T} - \frac{23474}{T} \right]$ & LTE & 11 \\
12 &  $\mH + \me  \rightarrow  \Hp + \me + \me $ &
$k_{12} =  \exp[-3.271396786 \times 10^{1}$ & & 12 \\
& & $\phantom{k_{12}=} \mbox{}  + 1.35365560 \times 10^{1} \ln T_{\rm e}$ & & \\
& & $\phantom{k_{12}=} \mbox{}  - 5.73932875 \times 10^{0} (\ln T_{\rm e})^{2}$  & & \\
& & $\phantom{k_{12}=} \mbox{}  + 1.56315498 \times 10^{0} (\ln T_{\rm e})^{3}$ & & \\
& & $\phantom{k_{12}=} \mbox{}  -  2.87705600 \times 10^{-1} (\ln T_{\rm e})^{4}$ & & \\
& & $\phantom{k_{12}=} \mbox{}  + 3.48255977 \times 10^{-2} (\ln T_{\rm e})^{5}$ & & \\
& & $\phantom{k_{12}=} \mbox{}   - 2.63197617 \times 10^{-3} (\ln T_{\rm e})^{6}$ & & \\ 
& & $\phantom{k_{12}=} \mbox{}  + 1.11954395\times 10^{-4} (\ln T_{\rm e})^{7}$ & & \\
& & $\phantom{k_{12}=} \mbox{}   -  2.03914985 \times 10^{-6} (\ln T_{\rm e})^{8}]$ & & \\
13 & $\Hp  + \me  \rightarrow \mH +  \gamma$ & 
 $k_{13, {\rm A}} = 1.269 \times 10^{-13} \left(\frac{315614}{T}\right)^{1.503}$ & Case A & 13 \\
& & $\phantom{k_{13, {\rm A}} = } \mbox{} \times [1.0+ \left(\frac{604625}{T}\right)^{0.470}]^{-1.923}$ & &  \\
& &  $k_{13, {\rm B}} = 2.753 \times 10^{-14} \left(\frac{315614}{T}\right)^{1.500}$ & Case B & 13 \\
& & $\phantom{k_{13, {\rm B}} = } \mbox{} \times [1.0+ \left(\frac{115188}{T}\right)^{0.407}]^{-2.242} $ & &  \\
14  &  $\Hm + \me  \rightarrow  \mH + \me + \me $ &
$ k_{14}  = \exp [-1.801849334 \times 10^{1}$ & & 12 \\
& & $\phantom{k_{14}=} \mbox{} + 2.36085220 \times 10^{0} \ln T_{\rm e}$ & &  \\
& & $\phantom{k_{14}=} \mbox{} - 2.82744300 \times 10^{-1} (\ln T_{\rm e})^{2}$ & & \\
& & $\phantom{k_{14}=} \mbox{}  +1.62331664\times 10^{-2} (\ln T_{\rm e})^{3}$ & & \\
& & $\phantom{k_{14}=} \mbox{} -3.36501203 \times 10^{-2} (\ln T_{\rm e})^{4}$ & & \\
& & $\phantom{k_{14}=} \mbox{}   +1.17832978\times 10^{-2} (\ln T_{\rm e})^{5}$ & & \\ 
& & $\phantom{k_{14}=} \mbox{}  -1.65619470\times 10^{-3} (\ln T_{\rm e})^{6}$ & & \\ 
& & $\phantom{k_{14}=} \mbox{}   +1.06827520\times 10^{-4} (\ln T_{\rm e})^{7}$ & & \\
& & $\phantom{k_{14}=} \mbox{}  -2.63128581\times 10^{-6} (\ln T_{\rm e})^{8} ]$ & & \\
\hline
\end{tabular}
\end{minipage}
\end{table*}

\begin{table*}
\begin{minipage}{126mm}
\contcaption{}
\begin{tabular}{llllc}
\hline 
No.\  & Reaction & Rate coefficient $({\rm cm}^{3} \: {\rm s}^{-1})$ & & Ref.\ \\
\hline
15  &  $\Hm + \mH  \rightarrow  \mH + \mH + \me $ &
$k_{15} = 2.5634 \times 10^{-9} T_{\rm e}^{1.78186}$ & $T_{\rm e} \leq 0.1 \: \rm{eV}$ & 12 \\
& & $\phantom{k_{15}} = \exp[-2.0372609 \times 10^{1}$ & $T_{\rm e} > 0.1 \: \rm{eV}$  & \\
& & $\phantom{k_{15} =} \mbox{}+1.13944933 \times 10^{0} \ln T_{\rm e}$ & & \\
& & $\phantom{k_{15} =} \mbox{}-1.4210135 \times 10^{-1} (\ln T_{\rm e})^{2}$ & & \\
& & $\phantom{k_{15} =} \mbox{}+8.4644554 \times 10^{-3} (\ln T_{\rm e})^{3}$ & & \\
& & $\phantom{k_{15} =} \mbox{}-1.4327641 \times 10^{-3} (\ln T_{\rm e})^{4}$  & & \\
& & $\phantom{k_{15} =} \mbox{}+2.0122503 \times 10^{-4} (\ln T_{\rm e})^{5}$ & & \\
& & $\phantom{k_{15} =} \mbox{}+8.6639632 \times 10^{-5} (\ln T_{\rm e})^{6}$ & & \\
& & $\phantom{k_{15} =} \mbox{}-2.5850097 \times 10^{-5} (\ln T_{\rm e})^{7}$ & & \\
& & $\phantom{k_{15} =} \mbox{}+ 2.4555012\times 10^{-6} (\ln T_{\rm e})^{8}$ & & \\
& & $\phantom{k_{15} =} \mbox{} -8.0683825\times 10^{-8} (\ln T_{\rm e})^{9}]$ & & \\
16  &  $\Hp + \Hm  \rightarrow  \mHtp + \me $ & 
$k_{16} =  6.9 \times 10^{-9}  T^{-0.35}$  & $T \le 8000 \: {\rm K}$ & 14 \\
& & $\phantom{k_{16}} =  9.6 \times 10^{-7} T^{-0.90}$ & $T > 8000 \: {\rm K}$ & \\
17 &  $\He + \me  \rightarrow  \Hep + \me + \me $ &
$k_{17} = \exp[-4.409864886 \times 10^{1}$ & & 12 \\
& & $\phantom{k_{17}=} \mbox{} + 2.391596563 \times 10^{1} \ln T_{\rm e}$ & & \\
& & $\phantom{k_{17}=} \mbox{} - 1.07532302 \times 10^{1} (\ln T_{\rm e})^{2}$ & & \\
& & $\phantom{k_{17}=} \mbox{} +3.05803875 \times 10^{0} (\ln T_{\rm e})^{3}$ & & \\
& & $\phantom{k_{17}=} \mbox{} - 5.68511890 \times 10^{-1} (\ln T_{\rm e})^{4}$ & & \\
& & $\phantom{k_{17}=} \mbox{} +6.79539123 \times 10^{-2} (\ln T_{\rm e})^{5}$ & & \\
& & $\phantom{k_{17}=} \mbox{} -5.00905610 \times 10^{-3} (\ln T_{\rm e})^{6}$ & & \\
& & $\phantom{k_{17}=} \mbox{} + 2.06723616\times 10^{-4}  (\ln T_{\rm e})^{7}$ & & \\
& & $\phantom{k_{17}=} \mbox{} - 3.64916141 \times 10^{-6} (\ln T_{\rm e})^{8}]$ & & \\ 
18 & $\Hep + \me \rightarrow \Hepp + \me + \me$ & 
$k_{18} = \exp[-6.87104099 \times 10^{1}$ & & 12 \\
& & $\phantom{k_{18}=} \mbox{} + 4.393347633 \times 10^{1} \ln T_{\rm e}$ & & \\
& & $\phantom{k_{18}=} \mbox{} - 1.84806699 \times 10^{1} (\ln T_{\rm e})^{2}$ & & \\
& & $\phantom{k_{18}=} \mbox{} + 4.70162649 \times 10^{0} (\ln T_{\rm e})^{3}$ & & \\
& & $\phantom{k_{18}=} \mbox{} - 7.6924663 \times 10^{-1} (\ln T_{\rm e})^{4}$ & & \\
& & $\phantom{k_{18}=} \mbox{} + 8.113042 \times 10^{-2} (\ln T_{\rm e})^{5}$ & & \\
& & $\phantom{k_{18}=} \mbox{} - 5.32402063 \times 10^{-3} (\ln T_{\rm e})^{6}$ & & \\
& & $\phantom{k_{18}=} \mbox{} + 1.97570531\times 10^{-4} (\ln T_{\rm e})^{7}$ & & \\
& & $\phantom{k_{18}=} \mbox{} - 3.16558106\times 10^{-6} (\ln T_{\rm e})^{8}$ & & \\
19 & $\Hep + \me \rightarrow \He + \gamma$ & 
$k_{\rm 19,  rr, A} = 10^{-11} T^{-0.5} \left[12.72 - 1.615 \log{T} \right. $ & Case A & 15  \\
& & $\left. \phantom{k_{\rm 19, rr, A} = } \mbox{} 
- 0.3162 (\log{T})^{2} + 0.0493 (\log{T})^{3}\right]$ & & \\
& & $k_{\rm 19, rr, B} = 10^{-11} T^{-0.5} \left[11.19 - 1.676 \log{T} \right. $ & Case B & 15  \\
& & $\left. \phantom{k_{\rm 19, rr, B} = } \mbox{} - 0.2852 (\log{T})^{2} 
+ 0.04433 (\log{T})^{3} \right]$ & & \\
& & $k_{\rm 19,  di} = 1.9 \times 10^{-3} T^{-1.5} \expf{-}{473421}{T}$ & Dielectronic & 16 \\
& & $\phantom{k_{19, {\rm di}} = } \mbox{} \times \left[1.0 + 0.3 \expf{-}{94684}{T} \right] $ & & \\
20 & $\Hepp + \me \rightarrow \Hep + \gamma$ & 
 $k_{20, {\rm A}} = 2.538 \times 10^{-13} \left(\frac{1262456}{T}\right)^{1.503}$ & Case A & 13 \\
& & $\phantom{k_{20, {\rm B}} = } \mbox{} \times 
 [1.0+ \left(\frac{2418500}{T}\right)^{0.470}]^{-1.923}$ & & \\
& & $k_{20, {\rm B}} = 5.506 \times 10^{-14} \left(\frac{1262456}{T}\right)^{1.500}$ & Case B & 13 \\
& & $\phantom{k_{20, {\rm B}} = } \mbox{} \times 
\left[1.0+ \left(\frac{460752}{T}\right)^{0.407}\right]^{-2.242} $ & & \\
21 & $\Hm + \mHtp \rightarrow \mHt + \mH$ & 
$k_{21} = 1.4 \times 10^{-7} \left(\frac{T}{300}\right)^{-0.5}$  & & 17 \\
22 & $\Hm + \mHtp \rightarrow \mH + \mH + \mH$ & 
$k_{22} = 1.4 \times 10^{-7} \left(\frac{T}{300}\right)^{-0.5}$  & & 17 \\
23 & $\mHt + \me \rightarrow \Hm + \mH$ & 
$k_{23} = 2.7 \times 10^{-8} T^{-1.27} \expf{-}{43000}{T}$ & & 18 \\
24 & $\mHt + \Hep \rightarrow \He + \mH + \Hp$ & 
$k_{24} = 3.7 \times 10^{-14} \expf{}{35}{T}$ & & 19 \\
25 & $\mHt + \Hep \rightarrow \mHtp + \He$ & $k_{25} = 7.2 \times 10^{-15}$ & & 19 \\
26 & $\Hep + \mH \rightarrow \He + \Hp$ & 
$k_{26} = 1.2 \times 10^{-15} \left(\frac{T}{300}\right)^{0.25}$ & & 20 \\
\hline
\end{tabular}
\end{minipage}
\end{table*}

\begin{table*}
\begin{minipage}{126mm}
\contcaption{}
\begin{tabular}{llllc}
\hline 
27 & $\He + \Hp \rightarrow \Hep + \mH$ &
$k_{27} = 1.26 \times 10^{-9} T^{-0.75} \expf{-}{127500}{T}$ & 
$ T \leq 10000 \: {\rm K}$ & 21 \\
& & $\phantom{k_{27}} = 4.0 \times 10^{-37} T^{4.74}$ & $T > 10000 \: {\rm K}$ & \\
28 & $\Hep + \Hm \rightarrow \He + \mH$ &
$k_{28} = 2.32 \times 10^{-7} \left(\frac{T}{300}\right)^{-0.52} \expf{}{T}{22400}$  & & 22 \\
29 & $\He + \Hm \rightarrow \He + \mH + \me$ & 
$k_{29} = 4.1 \times 10^{-17} T^{2} \expf{-}{19870}{T} $ & & 23 \\
30 & $\mH + \mH + \mH \rightarrow \mHt + \mH$ & See text  & & --- \\
31 & $\mH + \mH + \mHt \rightarrow \mHt + \mHt$ & See text & & --- \\
32 & $\mH + \mH + \He \rightarrow \mHt + \He$ & $k_{32} = 6.9 \times 10^{-32} T^{-0.4}$ & & 24 \\
33 & $\Dp + \me \rightarrow \mD + \gamma$ & $k_{33} = k_{13}$ & & 25 \\
34  &  $\mD + \Hp \rightarrow  \mH + \Dp $ & 
$k_{34} = 2.0 \times 10^{-10} T^{0.402}  \exp \left(-\frac{37.1}{T} \right)$ & 
$T\le 2 \times 10^{5} \: {\rm K}$ & 26 \\
& & $\phantom{k_{34}=} \mbox{}- 3.31 \times 10^{-17} T^{1.48}$ & & \\
& & $\phantom{k_{34}} = 3.44 \times 10^{-10} T^{0.35}$ & $T > 2 \times 10^{5} \: {\rm K}$ & \\
35 &  $\mH + \Dp  \rightarrow  \mD + \Hp $ & 
$k_{35} = 2.06 \times 10^{-10} T^{0.396}  \exp \left(-\frac{33}{T} \right)$ & & 26 \\
& & $\phantom{k_{35} =}\mbox{} + 2.03 \times 10^{-9} T^{-0.332}$ & & \\ 
36 & $\mH + \mD \rightarrow \hd + \gamma$ & 
$k_{36} = 10^{-25} [2.80202 - 6.63697 \ln T $ & $10 < T \le 200 \: {\rm K}$ & 27 \\
& & $\phantom{k_{36} = } \mbox{} + 4.75619 (\ln T)^{2} -1.39325 (\ln T)^{3} $ & & \\
& & $\phantom{k_{36}=}  \mbox{}  + 0.178259 (\ln T)^{4} - 0.00817097 (\ln T)^{5} ]$ & & \\
& & $\phantom{k_{36}} = 10^{-25} \exp [507.207 - 370.889 \ln T $ & $T > 200 \: {\rm K}$ & \\
& & $\phantom{k_{36}=}  \mbox{} + 104.854 (\ln T)^2 - 14.4192 (\ln T)^{3} $ & & \\
& & $\phantom{k_{36}=} \mbox{} + 0.971469 (\ln T)^{4} - 0.0258076 (\ln T)^{5} ]$ & & \\
37 & $\mHt + \mD \rightarrow \hd + \mH$ &
$k_{37} = {\rm dex}\left[-56.4737 + 5.88886\log{T} \right.$ & $T \le 2000 \: {\rm K}$ & 28 \\
& & $\phantom{k_{37} = {\rm dex}[} \mbox{} + 7.19692 (\log{T})^{2}$ & & \\
& & $\phantom{k_{37} = {\rm dex}[} \mbox{} + 2.25069 (\log{T})^{3}$ & & \\
& & $\phantom{k_{37} = {\rm dex}[} \mbox{} - 2.16903  (\log{T})^{4}$ & & \\
& & $\left. \phantom{k_{37} = {\rm dex}[} \mbox{} + 0.317887 (\log{T})^{5} \right]$ & & \\
& & $\phantom{k_{37}} = 3.17 \times 10^{-10} \expf{-}{5207}{T}$ & $T > 2000 \: {\rm K}$ & \\
38 & $\hdp + \mH \rightarrow \hd + \Hp$ & $k_{38} = k_{4}$ & & 25 \\
39 & $\mHt + \Dp \rightarrow \hd + \Hp$ & 
$k_{39} = \left[0.417 + 0.846 \log{T} - 0.137 (\log{T})^{2} \right] \times 10^{-9}$ & & 29 \\
40 &  $\hd + \mH  \rightarrow  \mHt + \mD $ &
$k_{40} = 5.25 \times 10^{-11} \expf{-}{4430}{T}$ & $T \le 200 \: {\rm K}$ & 30 \\
& & $\phantom{k_{40}} = 5.25 \times 10^{-11} \exp \left(-\frac{4430}{T} +
 \frac{173900}{T^{2}}\right)$ & $T > 200 \: {\rm K}$ & \\
41 & $\hd + \Hp \rightarrow \mHt + \Dp$ & $k_{41} = 1.1 \times 10^{-9} \expf{-}{488}{T}$ & & 29 \\
42 & $\mD + \Hp \rightarrow \hdp + \gamma$ & $k_{42} = 3.9 \times 10^{-19} 
\left(\frac{T}{300}\right)^{1.8} \expf{}{20}{T}$ & & 31 \\
43 & $\mH + \Dp \rightarrow \hdp + \gamma$ & $k_{43} = 3.9 \times 10^{-19} 
\left(\frac{T}{300}\right)^{1.8} \expf{}{20}{T}$ & & 31 \\
44 & $\hdp + \me \rightarrow \mH + \mD$ & $k_{44} = 7.2 \times 10^{-8} T^{-0.5}$ & & 32 \\
45 & $\mD + \me \rightarrow \Dp + \me + \me$ & $k_{45} =  k_{12}$ & & 25 \\
46 & $\Hep + \mD \rightarrow \Dp + \He$ & 
$k_{46} = 1.1 \times 10^{-15} \left(\frac{T}{300}\right)^{0.25}$ & & 31 \\
47 & $\He + \Dp \rightarrow \mD + \Hep$ &
$k_{47} = 1.85 \times 10^{-9} T^{-0.75} \expf{-}{127500}{T}$ & 
$ T \leq 10000 \: {\rm K}$ & 31 \\
& & $\phantom{k_{47}} = 5.9 \times 10^{-37} T^{4.74}$ & $T > 10000 \: {\rm K}$ & \\
48 & $\mHtp + \mD \rightarrow \hdp + \mH$ & 
$k_{48} = 1.07 \times 10^{-9} \left(\frac{T}{300}\right)^{0.062} \expf{-}{T}{41400}$ & & 33 \\
49 & $\hdp + \mD \rightarrow \hd + \Dp$  & $k_{49} = k_{4}$ & & 25 \\
50 & $\hdp + \mH \rightarrow \mHtp + \mD$  & $k_{50} = 1.0 \times 10^{-9} \expf{-}{154}{T}$ & & 34 \\
51  &  $\mD + \me  \rightarrow  \Dm + \gamma $ & $k_{51} = k_{1}$ & & 25 \\
52  &  $\mH + \Dm  \rightarrow  \mD + \Hm $ & 
$k_{52} =  6.4 \times 10^{-9} \left(\frac{T}{300}\right)^{0.41}$ & & 34 \\ 
53 &  $\mD + \Hm  \rightarrow  \mH + \Dm $ &
$k_{53} = 6.4 \times 10^{-9} \left(\frac{T}{300}\right)^{0.41}$ & & 34 \\ 
54  &  $\mD + \Hm  \rightarrow  \hd + \me $ & $k_{54} = 0.5 \, k_{2}$ & & 35 \\ 
55  &  $\mH + \Dm  \rightarrow  \hd + \me $ & $k_{55} = 0.5 \, k_{2}$ & & 35 \\
56 & $\mD + \Dm \rightarrow \DD + \me$ & $k_{56} = k_{2}$  & & 25 \\
57 &  $\hd + \me  \rightarrow  \mH + \Dm $ & 
$k_{57} = 1.35 \times 10^{-9} T^{-1.27} \expf{-}{43000}{T}$ & & 36 \\
58 &  $\hd + \me  \rightarrow  \mD + \Hm $ & 
$k_{58} = 1.35 \times 10^{-9} T^{-1.27} \expf{-}{43000}{T}$ & & 36 \\
59 &  $\DD + \me  \rightarrow  \mD + \Dm $ & 
$k_{59} = 6.7 \times 10^{-11} T^{-1.27} \expf{-}{43000}{T}$ & & 36 \\
60 & $\Hp + \Dm \rightarrow \hdp + \me$ & 
$k_{60} = 1.1 \times 10^{-9} \left(\frac{T}{300}\right)^{-0.4}$ & & 31 \\
61 & $\Dp + \Hm \rightarrow \hdp + \me$ & 
$k_{61} = 1.1 \times 10^{-9} \left(\frac{T}{300}\right)^{-0.4}$ & & 31 \\
62 & $\Dp + \Dm \rightarrow \ddp + \me$ & 
$k_{62} = 1.3 \times 10^{-9} \left(\frac{T}{300}\right)^{-0.4}$ & & 31 \\
\hline
\end{tabular}
\end{minipage}
\end{table*}

\begin{table*}
\begin{minipage}{126mm}
\contcaption{}
\begin{tabular}{llllc}
\hline 
63  &  $\Dm + \me  \rightarrow  \mD + \me + \me $ & $k_{63} = k_{14}$ & & 25 \\
64  &  $\Dm + \mH  \rightarrow  \mD + \mH + \me $ &  $k_{64} = k_{15}$ & & 25 \\
65 & $\Dm + \He  \rightarrow  \mD + \He + \me $ & 
$k_{65} = 1.5 \times 10^{-17} T^{2} \expf{-}{19870}{T}$ & & 31 \\
66 &  $\Dp + \Hm  \rightarrow  \mD + \mH $ & $k_{66} = k_{5}$ & & 25 \\
67  &  $\Hp + \Dm  \rightarrow  \mD + \mH $ & $k_{67} = k_{5}$  & & 25 \\
68  &  $\Dp + \Dm  \rightarrow  \mD + \mD $ & $k_{68} = k_{5}$  & & 25 \\
69 & $\mHtp + \Dm \rightarrow \mHt  + \mD$ & 
$k_{69} = 1.7 \times 10^{-7} \left(\frac{T}{300}\right)^{-0.5}$ & & 31 \\
70 & $\mHtp + \Dm \rightarrow \mH + \mH  + \mD$ & 
$k_{70} = 1.7 \times 10^{-7} \left(\frac{T}{300}\right)^{-0.5}$ & & 31 \\
71 & $\hdp + \Hm \rightarrow \hd  + \mH$ & 
$k_{71} = 1.5 \times 10^{-7} \left(\frac{T}{300}\right)^{-0.5}$ & & 31 \\
72 & $\hdp + \Hm \rightarrow \mD + \mH + \mH$ & 
$k_{72} = 1.5 \times 10^{-7} \left(\frac{T}{300}\right)^{-0.5}$ & & 31 \\
73 & $\hdp + \Dm \rightarrow \hd  + \mD$ & 
$k_{73} = 1.9 \times 10^{-7} \left(\frac{T}{300}\right)^{-0.5}$ & & 31 \\
74 & $\hdp + \Dm \rightarrow \mD + \mH + \mD$ & 
$k_{74} = 1.9 \times 10^{-7} \left(\frac{T}{300}\right)^{-0.5}$ & & 31 \\
75 & $\ddp + \Hm \rightarrow \DD + \mH$ & 
$k_{75} = 1.5 \times 10^{-7} \left(\frac{T}{300}\right)^{-0.5}$ & & 31 \\
76 & $\ddp + \Hm \rightarrow \mD + \mD + \mH$ & 
$k_{76} = 1.5 \times 10^{-7} \left(\frac{T}{300}\right)^{-0.5}$  & & 31 \\
77 & $\ddp + \Dm \rightarrow \DD + \mD$ & 
$k_{77} = 2.0 \times 10^{-7} \left(\frac{T}{300}\right)^{-0.5}$ & & 31 \\
78 & $\ddp + \Dm \rightarrow \mD + \mD + \mD$ & 
$k_{78} = 2.0 \times 10^{-7} \left(\frac{T}{300}\right)^{-0.5}$ & & 31 \\
79 &  $\Hep + \Dm  \rightarrow  \He + \mD $ & 
$k_{79} =  3.03 \times 10^{-7} \left(\frac{T}{300}\right)^{-0.52} \expf{}{T}{22400}$ & & 31 \\ 
80 & $\mD + \Dp \rightarrow \ddp + \gamma$ & 
$k_{80} =  1.9 \times 10^{-19} T_{3}^{1.8} \expf{}{20}{T}$ & & 31 \\
81  &  $\mD + \mHtp  \rightarrow  \mHt + \Dp $ & $k_{81} = k_{4}$ & & 25 \\ 
82 & $\mHtp + \mD \rightarrow \hd + \Hp$ & $k_{82} = 1.0 \times 10^{-9}$ & & 37 \\
83 & $\hdp + \mH \rightarrow \mHt + \Dp$ & $k_{83} = 1.0 \times 10^{-9}$ & & 37 \\
84 & $\hdp + \mD \rightarrow \ddp + \mH$ & $k_{84} = 1.0 \times 10^{-9}$ & & 38 \\
85 & $\hdp + \mD \rightarrow \DD + \Hp$ & $k_{85} = 1.0 \times 10^{-9}$ & & 37 \\
86  &  $\mD + \ddp  \rightarrow  \DD + \Dp $ & $k_{86} = k_{4}$ & & 25 \\
87  &  $\mH + \ddp  \rightarrow  \DD + \Hp $ & $k_{87} = k_{4}$ & & 25 \\ 
88 & $\ddp + \mH \rightarrow \hdp + \mD$ & $k_{88} = 1.0 \times 10^{-9} \expf{-}{472}{T}$ & & 38 \\
89 & $\ddp + \mH \rightarrow \hd + \Dp$ & $k_{89} = 1.0 \times 10^{-9}$  & & 37 \\
90  &  $\mHt + \Dp  \rightarrow  \mHtp + \mD $ & $k_{90} = k_{7}$ & & 25 \\ 
91 & $\mHt + \Dp \rightarrow \hdp + \mH$ & 
$k_{91} = \left[ 1.04 \times 10^{-9} + 9.52 \times 10^{-9} \left(\frac{T}{10000}\right) \right.$ & & 39 \\
& & $\phantom{k_{91}=} \left. \mbox{} -1.81 \times 10^{-9} \left(\frac{T}{10000}\right)^{2} 
\right]\expf{-}{21000}{T}$ & & \\
92 & $\hd + \Hp \rightarrow \hdp + \mH$ & $k_{92} = k_{7}$ & & 25 \\
93 & $\hd + \Hp  \rightarrow \mHtp + \mD$ &
$k_{93} = 1.0 \times 10^{-9} \expf{-}{21600}{T}$ & & 37 \\
94 & $\hd + \Dp \rightarrow \hdp + \mD$ & $k_{94} = k_{7}$ & & 25 \\
95 & $\hd + \Dp \rightarrow \DD + \Hp$ & $k_{95} = 1.0 \times 10^{-9}$ & & 38 \\
96 & $\hd + \Dp \rightarrow \ddp + \mH$ & 
$k_{96} = \left[ 3.54 \times 10^{-9} + 7.50 \times 10^{-10} \left(\frac{T}{10000}\right) \right.$ & & 39 \\
& & $\phantom{k_{96}=} \left. \mbox{} -2.92 \times 10^{-10} \left(\frac{T}{10000}\right)^{2} 
\right]\expf{-}{21100}{T}$ & & \\
97 & $\DD + \Hp \rightarrow \hd + \Dp$ & $k_{97} = 2.1 \times 10^{-9} \expf{-}{491}{T}$ & & 38 \\
98 & $\DD + \Hp \rightarrow \hdp + \mD$ & 
$k_{98} =  \left[ 5.18 \times 10^{-11} + 3.05 \times 10^{-9} \left(\frac{T}{10000}\right) \right.$ & & 39 \\
& & $\phantom{k_{98}=} \left. \mbox{} -5.42 \times 10^{-10} \left(\frac{T}{10000}\right)^{2} 
\right]\expf{-}{20100}{T}$ & & \\
99 & $\DD + \Hp \rightarrow \ddp + \mH$ & $k_{99} = k_{7}$ &  & 25 \\
100 & $\DD + \Dp \rightarrow \ddp + \mD$ & $k_{100} = k_{7}$ & & 25 \\
101 & $\hd + \Hep \rightarrow \hdp + \He$ & $k_{101} = k_{25}$ & & 25 \\
102  &  $\hd + \Hep  \rightarrow  \He + \Hp + \mD $ &
$k_{102} = 1.85 \times 10^{-14} \expf{}{35}{T}$ & & 35 \\
103  &  $\hd + \Hep  \rightarrow  \He + \mH + \Dp $ & 
$k_{103} = 1.85 \times 10^{-14} \expf{}{35}{T}$ & & 35 \\
104 & $\DD + \Hep \rightarrow \ddp + \He$ & $k_{104} = 2.5 \times 10^{-14}$ & & 38 \\
105  &  $\DD + \Hep  \rightarrow  \He + \Dp + \mD $ & 
$k_{105} = 1.1 \times 10^{-13} T_{3}^{-0.24}$ & & 38 \\
106 & $\hd + \mD \rightarrow \DD + \mH$ &  
$k_{106} = 1.15 \times 10^{-11} \expf{-}{3220}{T}$ & & 30 \\
\hline
\end{tabular}
\end{minipage}
\end{table*}

\begin{table*}
\begin{minipage}{126mm}
\contcaption{}
\begin{tabular}{llllc}
\hline 
107 & $\DD + \mH \rightarrow \hd + \mD$ & 
$k_{107} = {\rm dex}\left[-86.1558 + 4.53978 \log{T} \right.$ & $T \le 2200 \: {\rm K}$ &  28 \\
& & $\phantom{k_{107} = {\rm dex}[} \mbox{} + 33.5707 (\log{T})^{2}$ & & \\
& & $\phantom{k_{107} = {\rm dex}[} \mbox{} - 13.0449 (\log{T})^{3}$ & & \\
& & $\phantom{k_{107} = {\rm dex}[} \mbox{} + 1.22017  (\log{T})^{4}$ & & \\
& & $\left. \phantom{k_{107} = {\rm dex}[} \mbox{} + 0.0482453 (\log{T})^{5} \right]$ & & \\
& & $\phantom{k_{107}} = 2.67 \times 10^{-10} \expf{-}{5945}{T}$ & $T > 2200 \: {\rm K}$ & \\
108 & $\hd + \mH \rightarrow \mH + \mD + \mH$ & See text & & --- \\
109 & $\hd + \mHt \rightarrow \mH + \mD + \mHt$ & See text & & --- \\
110 & $\hd + \He \rightarrow \mH + \mD + \He$ & See text & & --- \\
111 & $\hd + \me \rightarrow \mH + \mD + \me$ & 
$k_{111} = 5.09 \times 10^{-9} T^{0.128}  \expf{-}{103258}{T}$ & $v=0$ & 40 \\
& & $\phantom{k_{111}} = 1.04 \times 10^{-9} T^{0.218}  \expf{-}{53070.7}{T}$ & LTE & \\
112 & $\DD + \mH \rightarrow \mD + \mD + \mH$ & $k_{112} = k_{9}$ & & 25 \\
113 & $\DD + \mHt \rightarrow \mD + \mD + \mHt$ & $k_{113} = k_{10}$ & & 25 \\
114 & $\DD + \He \rightarrow \mD + \mD + \He$ & $k_{114} = k_{11}$ & & 25 \\
115 & $\DD + \me \rightarrow \mD + \mD + \me$ & 
$k_{115} =  8.24 \times 10^{-9} T^{0.126} \expf{-}{105388}{T}$ & $v=0$ & 6 \\
& & $\phantom{k_{115}} = 2.75 \times 10^{-9} T^{0.163} \expf{-}{53339.7}{T}$ & LTE & \\
\hline
\end{tabular}
\medskip
\\
{\bf Note}: $T$ and $T_{\rm e}$ are the gas temperature in units of K 
and eV respectively. References are to the primary source of data for each 
reaction.\\
{\bf References}: 1: \citet{WIS79}, 2: \citet{RAM76}, 3: \citet{KAR79},
 4: \citet{SCH94}, 5: \citet{SAV04}, 6: \citet{tt02a},
 7: \citet{MAC86},  8:  \citet{ls83},  9: \citet{MAR98}, 10: \citet{sk87},
11: \citet{drcm87}, 12: \citet{JAN87}, 13: \citet{FER92}, 14: \citet{POU78},
15: \citet{hs98}, 16: \citet{ap73}, 17: \citet{dl87}, 18: \citet{sa67}, 19: \citet{b84},
20: \citet{z89}, 21: \citet{kldd93}, 22: \citet{ph94}, 23: \citet{h82}, 24: \citet{wk75},
25: Same as corresponding H reaction, 
26: \citet{sav02}, 27: \citet{dic05}, 28: Fit to data from \citet{mie03}, 
29: \citet{ger82}, 30: \citet{s59}, 31: Same as corresponding H reaction, 
but scaled by D reduced mass, 32: \citet{sss95}, 33: \citet{ljb95}, 
34: \citet{dm56}, scaled by D reduced mass,
35: Same as corresponding H reaction, with branching ratio assumed uniform,
36: \citet{xf01}, 37: estimate, 38: \citet{wfp04},  39: Fit based on cross-section from \citet{ws02}
40: \citet{tt02b}
\end{minipage}
\end{table*}

\end{document}